\documentclass[twocolumn,showpacs,preprintnumbers]{revtex4}
\usepackage{amssymb}
\usepackage{amsfonts}
\usepackage{amsmath}
\usepackage{graphicx}
\usepackage{dcolumn}
\usepackage{bm}

\input{tcilatex}
\begin{document}

\preprint{APS/123-QED}
\title{Optimization of the atom interferometer phase produced by the set of
cylindrical source masses to measure the Newtonian gravity constant}
\author{B. Dubetsky}
\email{bdubetsky@gmail.com}
\affiliation{1849 S Ocean Dr, Apt 207, Hallandale, FL 33009}
\date{\today}

\begin{abstract}
An analytical expression for the gravitational field of a homogeneous
cylinder is derived. The phase of the atom interferometer produced by the
gravity field of the set of cylinders has been calculated. The optimal
values of the initial positions and velocities of atomic clouds were
obtained. It is shown that at equal sizes of the atomic cloud in the
vertical and transverse directions, as well as at equal atomic vertical and
transverse temperatures, systematic errors due to the finite size and
temperature of the cloud disappear. To overcome the influence of the Earth
gravitational field on the accuracy of the phase double difference
measurement, it is proposed to use the technique of eliminating
gravity-gradient terms. After eliminating, one can use extreme values of the
atomic positions and velocities. Nonlinear dependences of the phase on the
uncertainties of atomic positions and velocities near those extreme values
required us to modify the expression for the standard phase deviation.
Moreover, such dependences lead to a phase shift, which was also calculated.
The relative accuracies of measurements of the Newtonian gravitational
constant $10^{-4}$ and $2\cdot 10^{-5}$ are predicted for sets of 24 and 630
cylinders, respectively.
\end{abstract}

\maketitle

\onecolumngrid%

\section{Introduction.}

Atom interferometry \cite{c1} is now used to measure Newtonian gravity
constant $G$ \cite{c2,c3,c4}. Searches for new schemes and options promise
to increase the accuracy of these measurements. Previously, it was shown 
\cite{c5} that, in principle, the current state-of-art in atom
interferometry would allow one to measure $G$ with an accuracy of $200$ppb$.$
To achieve such a goal one has to use \emph{simultaneously} the largest time
delay between pulses $T=1.15$ s \cite{c6}, temperature $115$pK and radius of
the atom cloud $170$ $\mu $ (which are larger than those observed in \cite%
{c7}), the beam splitter with an effective wave vector $k=8.25\cdot 10^{8}$m$%
^{\text{-1}}$ \cite{c8}, the source mass $1080$kg \cite{c9} and phase noise $%
\phi _{err}=10^{-4}$rad. For the parameters achieved in \cite{c2,c3,c4} at
present, the optimal preparation of the atomic clouds and proper positioning
of the gravity sources can also lead to an increase in the accuracy of the $%
G-$measurement.

The following procedure was\ used \cite{c3,c4,c5,c10}. The source mass
consists of two halves, which are placed in two different configurations C
and F. We accept the notation "C and F," which was previously used in
articles \cite{c3,c4}. The atomic gradiometer \cite{c4.1} measures the phase
difference of two atomic interferometers (AIs) $1$ and $2$ 
\begin{equation}
\Delta \phi ^{\left( C,F\right) }=\phi ^{\left( C,F\right) }\left(
z_{1},v_{z_{1}}\right) -\phi ^{\left( C,F\right) }\left(
z_{2},v_{z_{2}}\right) ,  \label{1}
\end{equation}%
where $\phi ^{\left( C,F\right) }\left( z_{j},v_{j}\right) $ is the phase of
AI $j$, in which the atoms are launched vertically from point $\mathbf{x}%
_{j}=\left( 0,0,z_{j}\right) $ at velocity $\mathbf{v}_{j}=\left(
0,0,v_{z_{j}}\right) $. Phase difference (\ref{1}) consists of two parts,
the one that is induced by the gravitational field of the Earth and inertial
terms and the one that is associated with the gravitational field of the
source mass. One expects \cite{c2,c3,c4} that the phase double difference
(PDD)%
\begin{equation}
\Delta ^{\left( 2\right) }\phi =\Delta \phi ^{\left( C\right) }-\Delta \phi
^{\left( F\right) }  \label{2}
\end{equation}%
will depend only on the AI phase $\phi _{s}^{\left( C,F\right) }\left(
z_{j},v_{j}\right) $ produced only by the field of source mass, and
therefore can be used to~measure the Newtonian gravitational constant $G$.
Despite the fact that the gravitational field of the Earth does not affect
the PDD, the gradient of this field affects \cite{c3,c4} on the accuracy of
the PDD measurement. In article \cite{c3,c4}, to reduce this influence, the
mutual position of the source mass and atomic clouds are selected so that at
the point of apogee of the atomic trajectories gradients of the Earth's
field and the field of the source mass cancel each other. Below in Sec. \ref%
{s4} we will see that this technique only partially reduces the influence of
the gravitational field of the Earth on the accuracy of the $G$ measurement.

To resolve this problem we suggest using the method of eliminating the
sensitivity of the phase to the atom position and velocity \cite{c17}, which
is achieved by changing the wave vector of the second Raman pulse, so that%
\begin{equation}
\mathbf{k}_{2}=\mathbf{k}+\dfrac{1}{2}\Gamma _{E}T^{2}\mathbf{k},
\label{2.9}
\end{equation}%
where $\Gamma _{E}$ is gravity-gradient tensor of the Earth field. The
possibility of using the technique of \cite{c17} for measuring $G$ was
considered by G. Rosi \cite{c15}. The difference from what is proposed here
is that we propose to eliminate only the gradient of the gravitational field
of the Earth. Since the gravity gradient tensor is measured with some
accuracy $\delta \Gamma _{E},$ it still contributes to the error budget. One
can ignore this influence if the relative standard deviation (RSD) of the
PDD, $\sigma _{r}\left( \Delta ^{\left( 2\right) }\phi \right) ,$ is
sufficiently large,

\begin{equation}
\sigma _{r}\left( \Delta ^{\left( 2\right) }\phi \right) >\dfrac{k\left\vert
\delta \Gamma _{E}\right\vert T^{2}}{\Delta ^{\left( 2\right) }\phi }\left\{
\dsum_{j=1,2}\dsum_{I=C,F}\left[ \sigma ^{2}\left( z_{jI}\right)
+T^{2}\sigma ^{2}\left( v_{z_{jI}}\right) \right] \right\} ^{1/2},
\label{2.10}
\end{equation}%
where $\sigma \left( z_{jI}\right) $ and $\sigma \left( v_{z_{jI}}\right) $
are the standard deviations (SD) of the initial atomic position and
velocity. Although, starting with article \cite{c4.1}, methods for measuring
the gravity gradient using AIs have been studied in many articles, I know
only three publications \cite{c1.4,c1.5,c1.6}, in which the values of $%
\Gamma _{E33}$ and $\delta \Gamma _{E33}$ were published. The error $\delta
\Gamma _{E33}=10E$ was reported \cite{c1.6}. For the values obtained in \cite%
{c3,c4}, $\sigma \left( z_{jI}\right) \thicksim 10^{-4}$m, $\sigma \left(
v_{z_{jI}}\right) \thicksim 3\cdot 10^{-3}$m/s, $k\approx 1.61\cdot 10^{7}$m$%
^{\text{-1}}$,%
\begin{equation}
\Delta ^{\left( 2\right) }\phi =0.547870(63)\text{rad,}  \label{2.7}
\end{equation}%
one gets%
\begin{equation}
\sigma _{r}\left( \Delta ^{\left( 2\right) }\phi \right) >7\text{ppm}.
\label{2.11}
\end{equation}
We will make sure that the restriction (\ref{2.11}) can be neglected with an
accuracy of no more than 10\%.

Since the magnitude of the gravity-gradient tensor $\Gamma _{E}$ is small,
the change in the effective wave vector in (\ref{2.9}) can be considered as
a perturbation. Another small perturbation here is the gravity field of the
source mass \cite{c10}. Since we are not going to consider the simultaneous
action of these two perturbations, we can calculate the parts of the phases $%
\phi _{s}^{\left( C,F\right) }$ assuming that all three Raman pulses have
the same unperturbed effective wave vector $\mathbf{k.}$

In order to achieve the maximum value of the first order difference $\Delta
\phi _{s}^{\left( C\right) }$ in $C-$configuration, one can choose points $%
\left\{ z_{1},v_{z_{1}}\right\} $ and $\left\{ z_{2},v_{z_{2}}\right\} $ at
the maximum and minimum of the dependence $\phi _{s}^{\left( C\right)
}\left( z,v\right) $ \cite{c5,c10}. The choice of extreme points leads to a
quadratic dependence of the phase $\phi _{s}^{\left( C\right) }\left(
z,v\right) $ on small deviations $\left\{ \delta z_{j},\delta v_{j}\right\} $
in the vicinities of these points.

Let us consider now the contribution to the phase double difference from the
first-order phase difference in $F-$configuration. In principle, one can
find the positions of the halves of the source mass, at which points $z_{1}$
and $z_{2}$ are extrema of the dependence $\phi _{s}^{\left( F\right)
}\left( z,v\right) $ \cite{c12}. However, in this case, points in the
velocity space $v_{z_{1}}$ and $v_{z_{2}},$ which were extreme for the $C-$%
configuration, become non-extreme. To avoid this difficulty one can \cite%
{c5,c10} distance the halves of the source mass sufficiently far so that
even the linear dependences of the $\phi _{s}^{\left( F\right) }\left(
z,v\right) $ on the deviations near the points $\left\{ z_{j},v_{j}\right\} $
do not affect significantly the error $\phi _{s}^{\left( C\right) }$\ of the
phase double differences (\ref{2}).

We performed \cite{c5,c10} calculations, determined the optimal geometry of
the gravitational field, positions and velocities of atomic clouds for the
source mass of a cuboid shape. The choice of this shape is convenient for
calculations since one has an analytical expression for the potential of the
cuboid \cite{c13}. Despite this, it is preferable to use the source mass in
a cylindrical shape to perform high-precision measurements of $G$ \cite{c14}%
. Cylindrical source masses were used to measure $G$ with an accuracy of $%
150 $ppm \cite{c3,c4}. The hollow cylinder source mass has been proposed to
achieve an accuracy of 10ppm \cite{c15}. The analytical expression for the
gravitational field along the $z-$axis of the hollow cylinder was explored 
\cite{c15}, but outside this axis, the potential expansion into spherical
harmonics was used \cite{c3,c4}. Expressions for the field of the cylinders
have been derived in the articles \cite{c15.1,c15.2}. Alternatively, the
technique for calculating the gravitational field without calculating the
gravitational potential was proposed in the book \cite{c14}, but the final
expression for the cylinder field is given in \cite{c14} without derivation.
Following technique \cite{c14}, we calculated the field and arrived at
expressions (\ref{a16}, \ref{a20}). Our expressions do not coincide with
those given in \cite{c14,c15.1,c15.2}. Both the derivations and final
results are presented in this article. Following the derivations in the
articles \cite{c15.1,c15.2}, we are going to find out analytically the
reason of the discrepancies between different expressions and publish it
elsewhere.

To estimate the expected accuracy of the $G-$measurement, one has to analyze
the error $\phi _{s}^{\left( C\right) }$. In precision gravity experiments,
one calculates or measures the SD $\sigma $ of the response $f$ (such as the
AI phase or phase difference) using the expression

\begin{equation}
\sigma \left( f\right) =\left( \sum_{m=1}^{n}\sigma _{m}^{2}\right) ^{1/2},
\label{3}
\end{equation}%
where $n$ is the number of variables $\left\{ q_{1},\ldots q_{n}\right\} $,
included in the error budget, $\sigma _{m}=\left\vert \partial f/\partial
q_{m}\right\vert \sigma \left( q_{m}\right) ,$ and.$\sigma \left(
q_{m}\right) $ is a SD of small uncertainty in the variable $q_{m}.$ We
assume that variables $\left\{ q_{1},\ldots q_{n}\right\} $ are
statistically independent. See examples of such budgets in \cite%
{c2,c3,c4,c1.4,c12,c15,c16}. The situation changes when one considers
uncertainties near the extreme points $\left\{ \mathbf{x}_{m},\mathbf{v}%
_{m}\right\} $ and the signal's uncertainty becomes a quadratic function of
the uncertainties of the atomic position and velocity $\left\{ \delta 
\mathbf{x}_{m},\delta \mathbf{v}_{m}\right\} .$ There are several examples
in which measurements were carried out (or proposed to be carried out) near
extreme points. Extreme atomic coordinates were selected in the experiments 
\cite{c12}. Extreme atomic coordinates and velocities were found in the
articles \cite{c5,c10}. The difficulties of using extreme points are noted
in the article \cite{c15}, where an alternative approach was proposed, based
on the elimination of the dependence of the AI phase on the atomic position
and velocity proposed in \cite{c17}. However, even in this case, one
eliminates only the dependence on the vertical coordinates and velocities,
while the transverse coordinates $\left\{ x_{m},y_{m}\right\} =\left\{
0,0\right\} $ and velocities $\left\{ v_{x_{m}},v_{y_{m}}\right\} =\left\{
0,0\right\} $ remain extreme. This is because the vertical component of the
gravitational field of the hollow cylinder $\delta g_{3}\left( \mathbf{x}%
\right) $ is axially symmetric, and the expansion of both the field and the
field gradient in transverse coordinates begins with quadratic terms. We see
that in all the cases listed above \cite{c5,c10,c12,c15}, the use of the
expression (\ref{3}) is unjustified. Revision of this expression is
required. Moreover, the quadratic dependence on the uncertainties $\left\{
\delta \mathbf{x}_{m},\delta \mathbf{v}_{m}\right\} $ leads to a shift in
the signal \cite{c18}. Here, we expressed both the SD and the shift of the
phase double difference (\ref{2}) in terms of the first and second
derivatives of the phases $\phi _{s}^{\left( C,F\right) }$ at the found
extreme points.

The article is arranged as follows.\ Standard deviation and shift are
obtained in the next section. Section \ref{s3} is devoted to the AI phase
and phase derivatives calculations, PDD and error budget for source mass
consisting of 24 cylinders are obtained in\ the Sec. \ref{s4}. An
optimization procedure in respect to atomic positions and velocities
considered in\ the Sec. V. The case of the 630 cylinders source mass is
studied in the Sec. \ref{s6}, while the derivation of the cylinder
gravitational field is presented in the Appendix.

\section{\label{s2}SD and shift.}

Let us consider the variation of the double difference (\ref{2})%
\begin{gather}
\delta \Delta ^{\left( 2\right) }\phi \left[ \delta \mathbf{x}_{1C},\delta 
\mathbf{v}_{1C},\delta \mathbf{x}_{2C},\delta \mathbf{v}_{2C};\delta \mathbf{%
x}_{1F},\delta \mathbf{v}_{1F};\delta \mathbf{x}_{2F},\delta \mathbf{v}_{2F}%
\right]  \notag \\
=\delta \phi ^{\left( C\right) }\left[ \delta \mathbf{x}_{1C},\delta \mathbf{%
v}_{1C}\right] -\delta \phi ^{\left( C\right) }\left[ \delta \mathbf{x}%
_{2C},\delta \mathbf{v}_{2C}\right]  \notag \\
-\left[ \delta \phi ^{\left( F\right) }\left( \delta \mathbf{x}_{1F},\delta 
\mathbf{v}_{1F}\right) -\delta \phi ^{\left( F\right) }\left( \delta \mathbf{%
x}_{2F},\delta \mathbf{v}_{2F}\right) \right] ,  \label{4}
\end{gather}%
where $\left\{ \delta \mathbf{x}_{jI},\delta \mathbf{v}_{jI}\right\} $ is
the uncertainty of the launching position and velocity of the cloud $j$ $%
\left( j=1\text{ or }2\right) $ for the source mass configuration $I$ $%
\left( I=C\text{ or }F\right) ,~\delta \phi ^{\left( I\right) }\left( \delta 
\mathbf{x}_{jI},\delta \mathbf{v}_{jI}\right) $ is the variation of the AI $%
j $ phase, produced when the source mass gravity field is in the $I-$%
configuration. For the shift $s$ and standard deviation $\sigma $ defined as 
\begin{subequations}
\label{5}
\begin{eqnarray}
s\left( \Delta ^{\left( 2\right) }\phi \right) &=&\left\langle \delta \Delta
^{\left( 2\right) }\phi \left[ \delta \mathbf{x}_{1C},\delta \mathbf{v}%
_{1C},\delta \mathbf{x}_{2C},\delta \mathbf{v}_{2C};\delta \mathbf{x}%
_{1F},\delta \mathbf{v}_{1F};\delta \mathbf{x}_{2F},\delta \mathbf{v}_{2F}%
\right] \right\rangle ,  \label{5a} \\
\sigma \left( \Delta _{s}^{\left( 2\right) }\phi \right) &=&\left\{
\left\langle \left[ \delta \Delta ^{\left( 2\right) }\phi \left( \delta 
\mathbf{x}_{1C},\delta \mathbf{v}_{1C},\delta \mathbf{x}_{2C},\delta \mathbf{%
v}_{2C};\delta \mathbf{x}_{1F},\delta \mathbf{v}_{1F};\delta \mathbf{x}%
_{2F},\delta \mathbf{v}_{2F}\right) \right] ^{2}\right\rangle -s^{2}\left(
\Delta ^{\left( 2\right) }\phi \right) \right\} ^{1/2}  \label{5b}
\end{eqnarray}%
one finds 
\end{subequations}
\begin{subequations}
\label{6}
\begin{gather}
s\left( \Delta ^{\left( 2\right) }\phi \right) =s\left[ \phi ^{\left(
C\right) }\left( \delta \mathbf{x}_{1C},\delta \mathbf{v}_{1C}\right) \right]
-s\left[ \phi ^{\left( C\right) }\left( \delta \mathbf{x}_{2C},\delta 
\mathbf{v}_{2C}\right) \right] -s\left[ \phi ^{\left( F\right) }\left(
\delta \mathbf{x}_{1F},\delta \mathbf{v}_{1F}\right) \right] +s\left[ \phi
^{\left( F\right) }\left( \delta \mathbf{x}_{2F},\delta \mathbf{v}%
_{2F}\right) \right] ,  \label{6a} \\
\sigma \left( \Delta ^{\left( 2\right) }\phi \right) =\left\{
\dsum_{I=C,F}\dsum_{j=1,2}\sigma ^{2}\left[ \phi ^{\left( I\right) }\left(
\delta \mathbf{x}_{jI},\delta \mathbf{v}_{jI}\right) \right] \right\} ^{1/2},
\label{6b} \\
s\left[ \phi ^{\left( I\right) }\left( \delta \mathbf{x}_{jI},\delta \mathbf{%
v}_{jI}\right) \right] =\left\langle \delta \phi ^{\left( I\right) }\left(
\delta \mathbf{x}_{jI},\delta \mathbf{v}_{jI}\right) \right\rangle
\label{6c} \\
\sigma \left[ \phi ^{\left( I\right) }\left( \delta \mathbf{x}_{jI},\delta 
\mathbf{v}_{jI}\right) \right] =\left\{ \left\langle \left[ \delta \phi
^{\left( I\right) }\left( \delta \mathbf{x}_{jI},\delta \mathbf{v}%
_{jI}\right) \right] ^{2}\right\rangle -\left\langle \delta \phi ^{\left(
I\right) }\left( \delta \mathbf{x}_{jI},\delta \mathbf{v}_{jI}\right)
\right\rangle ^{2}\right\} ^{1/2}  \label{6d}
\end{gather}%
One sees that the problem is reduced to the calculation of the shift $s$ and
SD $\sigma $ of a variation $\delta \phi \left[ \delta \mathbf{x},\delta 
\mathbf{v}\right] .$ The phase of the given AI at the given configuration of
the source mass comprises two parts 
\end{subequations}
\begin{equation}
\phi \left( \mathbf{x},\mathbf{v}\right) =\phi _{E}\left( \mathbf{x},\mathbf{%
v}\right) +\phi _{s}\left( \mathbf{x},\mathbf{v}\right) ,  \label{2.3}
\end{equation}%
where for the phase induced by the Earth's field, under some simplifying
assumptions (see, for example, \cite{c19}), one gets 
\begin{equation}
\phi _{E}^{\left( I\right) }\left( \mathbf{x}_{j},\mathbf{v}_{j}\right) =%
\mathbf{k\cdot g}T^{2}+\mathbf{k\cdot }\Gamma _{E}T^{2}\left( \mathbf{x}+%
\mathbf{v}T\right) +\mathbf{k\cdot }\Gamma _{E}\mathbf{g}T^{2}\left( \dfrac{7%
}{12}T^{2}+TT_{1}+\dfrac{1}{2}T_{1}^{2}\right) ,  \label{2.4}
\end{equation}%
where $T_{1}$ is the time delay between the moment the atoms are launched
and the 1st Raman pulse. For the vertical wave vector $\mathbf{k=}\left(
0,0,k\right) ,$ expanding Eq. (\ref{2.3}).to the second order terms one gets 
\begin{equation}
\delta \phi \left( \delta \mathbf{x},\delta \mathbf{v}\right) =\left( \tilde{%
\gamma}_{xm}+\dfrac{\partial \phi _{s}}{\partial x_{m}}\right) \delta
x_{m}+\left( \tilde{\gamma}_{vm}+\dfrac{\partial \phi _{s}}{\partial v_{m}}%
\right) \delta v_{m}+\dfrac{1}{2}\dfrac{\partial ^{2}\phi _{s}}{\partial
x_{m}\partial x_{n}}\delta x_{m}\delta x_{n}+\dfrac{1}{2}\dfrac{\partial
^{2}\phi _{s}}{\partial v_{m}\partial v_{n}}\delta v_{m}\delta v_{n}+\dfrac{%
\partial ^{2}\phi _{s}}{\partial x_{m}\partial v_{n}}\delta x_{m}\delta
v_{n},  \label{7}
\end{equation}%
where 
\begin{subequations}
\label{7.1}
\begin{eqnarray}
\tilde{\gamma}_{xm} &=&k\Gamma _{E3m}T^{2};  \label{7.1a} \\
\tilde{\gamma}_{vm} &=&T\tilde{\gamma}_{xm}  \label{7.1b}
\end{eqnarray}%
A summation convention implicit in Eq. (\ref{7}) will be used in all
subsequent equations. Repeated indices and symbols appearing on the
right-hand-side (rhs) of an equation are to be summed over, unless they also
appear on the left-hand-side (lhs) of that equation. Let assume that the
distribution functions of the uncertainties are sufficiently symmetric, and
all odd moments are equal $0.$ The moments of the second and fourth orders
are given by 
\end{subequations}
\begin{subequations}
\label{8}
\begin{eqnarray}
\left\langle \delta q_{m}\delta q_{n}\right\rangle &=&\delta _{mn}\sigma
^{2}\left( q_{m}\right) ,  \label{8a} \\
\left\langle \delta q_{m}\delta q_{n}\delta q_{m^{\prime }}\delta
q_{n^{\prime }}\right\rangle &=&\delta _{mn}\delta _{m^{\prime }n^{\prime
}}\sigma ^{2}\left( q_{m}\right) \sigma ^{2}\left( q_{m^{\prime }}\right)
+\left( \delta _{mm^{\prime }}\delta _{nn^{\prime }}+\delta _{mn^{\prime
}}\delta _{nm^{\prime }}\right) \sigma ^{2}\left( q_{m}\right) \sigma
^{2}\left( q_{n}\right)  \notag \\
&&+\delta _{mn}\delta _{mm^{\prime }}\delta _{mn^{\prime }}\kappa \left(
q_{m}\right) \sigma ^{4}\left( q_{m}\right) ,  \label{8b}
\end{eqnarray}%
where $q_{i}$ is either a position $x_{i}$ or a velocity $v_{i},$ $\sigma
\left( q_{i}\right) $ is SD of the uncertainty $\delta q_{i},$ $\delta _{mn}$
is Kronecker symbol, $\kappa \left( q_{m}\right) $ is a cumulant of the
given uncertainty $\delta q_{m}$, defined as 
\end{subequations}
\begin{equation}
\kappa \left( q_{m}\right) =\dfrac{\left\langle \delta
q_{m}^{4}\right\rangle }{\sigma ^{4}\left( q_{m}\right) }-3.  \label{9}
\end{equation}%
Using the moments (\ref{8}) one arrives at the following expressions for the
SD and shift 
\begin{subequations}
\label{10}
\begin{eqnarray}
\sigma \left[ \phi \left( \delta \mathbf{x},\delta \mathbf{v}\right) \right]
&=&\left\{ \left( \tilde{\gamma}_{xm}+\dfrac{\partial \phi _{s}}{\partial
x_{m}}\right) ^{2}\sigma ^{2}\left( x_{m}\right) +\left( \tilde{\gamma}_{vm}+%
\dfrac{\partial \phi _{s}}{\partial v_{m}}\right) ^{2}\sigma ^{2}\left(
v_{m}\right) \right.  \notag \\
&&+\dfrac{1}{2}\left[ \left( \dfrac{\partial ^{2}\phi }{\partial
x_{m}\partial x_{n}}\right) ^{2}\sigma ^{2}\left( x_{m}\right) \sigma
^{2}\left( x_{n}\right) +\left( \dfrac{\partial ^{2}\phi }{\partial
v_{m}\partial v_{n}}\right) ^{2}\sigma ^{2}\left( v_{m}\right) \sigma
^{2}\left( v_{n}\right) \right]  \notag \\
&&\left. +\left( \dfrac{\partial ^{2}\phi }{\partial x_{m}\partial v_{n}}%
\right) ^{2}\sigma ^{2}\left( x_{m}\right) \sigma ^{2}\left( v_{n}\right) +%
\dfrac{1}{4}\left[ \left( \dfrac{\partial ^{2}\phi }{\partial x_{m}^{2}}%
\right) ^{2}\kappa \left( x_{m}\right) \sigma ^{4}\left( x_{m}\right)
+\left( \dfrac{\partial ^{2}\phi }{\partial v_{m}^{2}}\right) ^{2}\kappa
\left( v_{m}\right) \sigma ^{4}\left( v_{m}\right) \right] \right\} ^{1/2},
\label{10a} \\
s\left[ \phi \left( \delta \mathbf{x},\delta \mathbf{v}\right) \right] &=&%
\dfrac{1}{2}\left( \dfrac{\partial ^{2}\phi }{\partial x_{m}^{2}}\sigma
^{2}\left( x_{m}\right) +\dfrac{\partial ^{2}\phi }{\partial v_{m}^{2}}%
\sigma ^{2}\left( v_{m}\right) \right) ,  \label{10b}
\end{eqnarray}%
One sees that, even for the symmetric uncertainties distribution, the
knowledge of the uncertainties' SDs is not sufficient. One has to know also
uncertainties' cumulants (\ref{9}). The exclusion here is Gaussian
distributions, for which the cumulants 
\end{subequations}
\begin{equation}
\kappa \left( x_{m}\right) =\kappa \left( v_{m}\right) =0.  \label{10.1}
\end{equation}%
Further calculations will be performed only for these distributions.

For the each case considered below we are going to calculate the double
difference (\ref{2})$\ $and relative contributions to the shift (\ref{5a})
and SD (\ref{5b}) from the each of two atom clouds at the each of two source
mass configurations.

\section{\label{s3}The phase and phase derivatives of the atom interferometer%
}

To calculate the phase $\phi _{s}^{\left( I\right) }$ produced by the
gravitational field of the source mass, we use the results obtained in the
article \cite{c20}.\ It is necessary to distinguish three contributions to
the phase, classical, quantum, and Q-term (see Eqs. (62c, 64, 60c), (62d,
71, 60c), (89) in \cite{c20} for these three terms). For Q-term an estimate
was obtained 
\begin{equation}
\dfrac{\phi _{Q}}{\phi _{s}^{\left( I\right) }}\thicksim \dfrac{1}{24}\left( 
\dfrac{\hbar kT}{LM_{a}}\right) ^{2},  \label{11}
\end{equation}%
where $M_{a}$ is the atom mass, $L$ is the characteristic distance over
which the gravitational potential of the test mass changes. For $^{\text{87}%
} $Rb, at $L>0.3$m, the relative weight of the Q-term does not exceed 2ppb,
and we neglect it. For the remaining terms and the vertical effective wave
vector, $\mathbf{k}=\left\{ 0,0,k\right\} ,$ one gets%
\begin{equation}
\phi _{s}^{\left( I\right) }\left( \mathbf{x},\mathbf{v}\right)
=k\int_{0}^{T}dt\left[ \left( T-t\right) \delta g_{3}\left( \mathbf{a}\left(
T+t\right) \right) +t\delta g_{3}\left( \mathbf{a}\left( t\right) \right) %
\right] ,  \label{12}
\end{equation}%
where%
\begin{equation}
\mathbf{a}\left( t\right) =\mathbf{x}+\mathbf{v}\left( T_{1}+t\right) +%
\dfrac{1}{2}\mathbf{g}\left( T_{1}+t\right) ^{2}+\mathbf{v}_{r}t,
\label{12.1}
\end{equation}%
the recoil velocity is given by%
\begin{equation}
\mathbf{v}_{r}=\hbar \mathbf{k/}2M_{a},  \label{12.2}
\end{equation}%
$\delta g_{3}\left( \mathbf{x}\right) $ is the vertical component of the
gravitational field of the source mass. The derivatives of this phase of the
first and second order are given by 
\begin{subequations}
\label{13}
\begin{eqnarray}
\dfrac{\partial \phi _{s}^{\left( I\right) }\left( \mathbf{x},\mathbf{v}%
\right) }{\partial x_{m}} &=&k\int_{0}^{T}dt\left[ \left( T-t\right) \Gamma
_{s3m}\left( \mathbf{a}\left( T+t\right) \right) +t\Gamma _{s3m}\left( 
\mathbf{a}\left( t\right) \right) \right] ,  \label{13a} \\
\dfrac{\partial \phi _{s}^{\left( I\right) }\left( \mathbf{x},\mathbf{v}%
\right) }{\partial v_{m}} &=&k\int_{0}^{T}dt\left[ \left( T-t\right) \left(
T_{1}+T+t\right) \Gamma _{s3m}\left( \mathbf{a}\left( T+t\right) \right)
+t\left( T_{1}+t\right) \Gamma _{s3m}\left( \mathbf{a}\left( t\right)
\right) \right] ,  \label{13b} \\
\dfrac{\partial ^{2}\phi _{s}^{\left( I\right) }\left( \mathbf{x},\mathbf{v}%
\right) }{\partial x_{m}\partial x_{n}} &=&k\int_{0}^{T}dt\left[ \left(
T-t\right) \chi _{s3mn}\left( \mathbf{a}\left( T+t\right) \right) +t\chi
_{s3mn}\left( \mathbf{a}\left( t\right) \right) \right] ,  \label{13c} \\
\dfrac{\partial ^{2}\phi _{s}^{\left( I\right) }\left( \mathbf{x},\mathbf{v}%
\right) }{\partial x_{m}\partial v_{n}} &=&k\int_{0}^{T}dt\left\{ \left(
T-t\right) \left( T_{1}+T+t\right) \chi _{s3mn}\left( \mathbf{a}\left(
T+t\right) \right) +t\left( T_{1}+t\right) \chi _{s3mn}\left( \mathbf{a}%
\left( t\right) \right) \right. ,  \label{13d} \\
\dfrac{\partial ^{2}\phi _{s}^{\left( I\right) }\left( \mathbf{x},\mathbf{v}%
\right) }{\partial v_{m}\partial v_{n}} &=&k\int_{0}^{T}dt\left[ \left(
T-t\right) \left( T_{1}+T+t\right) ^{2}\chi _{s3mn}\left( \mathbf{a}\left(
T+t\right) \right) +t\left( T_{1}+t\right) ^{2}\chi _{s3mn}\left( \mathbf{a}%
\left( t\right) \right) \right] ,  \label{13e}
\end{eqnarray}%
where $\Gamma _{s3m}\left( \mathbf{x}\right) =\dfrac{\partial \delta
g_{3}\left( \mathbf{x}\right) }{\partial x_{m}}$ is the $3m-$component of
the gravity-gradient tensor of the source mass field, and 
\end{subequations}
\begin{equation}
\chi _{s3mn}\left( \mathbf{x}\right) =\dfrac{\partial ^{2}\delta g_{3}\left( 
\mathbf{x}\right) }{\partial x_{m}\partial x_{n}}  \label{13.1}
\end{equation}%
is the $3mn-$component of the curvature tensor of this field.

\section{\label{s4}Error budget}

\begin{figure}[!t]
\includegraphics[width=15cm]{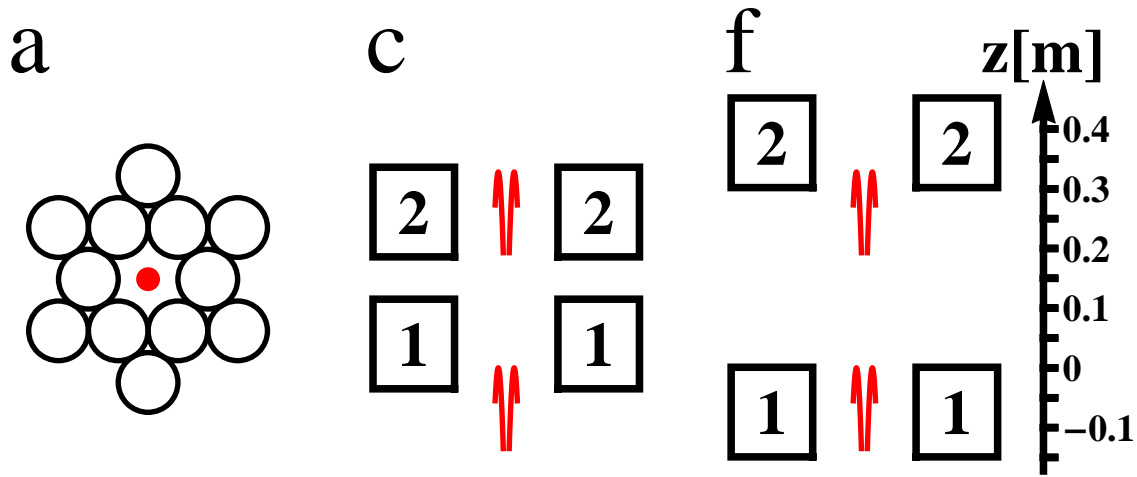}
\caption{The mutual positioning of the
source mass halves 1 and 2, and atomic clouds. Top view (a), cross-sections
x=0 for C-configuration (c) and F-configuration (f). Trajectories of atoms
are shown in red.}
\label{f1}
\end{figure}

We applied the formula for the cylinder field (\ref{a16}) to calculate the
phases produced by different sets of cylinders. In this section, we consider
the field geometry chosen in the article \cite{c3,c4}, see Fig. \ref{f1}.

Two halves of the source mass, each including 12 tungsten alloy cylinders,
move in a vertical direction from $C-$configuration to $F-$configuration, in
each of which one measures the phase difference of the first order (\ref{1}%
), and then PDD (\ref{2}). The following system parameters are important for
calculation: cylinder density $\rho =18263$kg/m%
${{}^3}$%
, cylinder radius and height $R=0.0495$m and $h=0.15011$m, Newtonian
gravitational constant $G=6.67408\cdot 10^{-11}$kg$^{\text{-1}}$m$^{\text{3}%
} $s$^{\text{-2}}$ \cite{c21}, the Earth's gravitational field $g=9.80492$m/s%
$^{\text{2}}$ \cite{c22}, the delay between impulses $T=160$ms, the time $%
T_{1}=0,$ the effective wave vector $k=1.61058\cdot 10^{7}$m$^{\text{-1}}$,
the mass of the $^{\text{87}}$Rb $M_{a}=86.9092$a.u. \cite{c23}, atomic
velocity at the moment of the first impulse action $v=1.62762$m/s. With
respect to the apogee of the atomic trajectory in the lower interferometer,
the $z-$coordinates of the centers of the halves of the source mass are
equal to $0.04$m and $0.261$m in the $C-$configuration and $-0.074m$ and $%
0.377$m in the $F-$configuration, $z-$coordinate of the atomic trajectory
apogee in the upper interferometer is equal to $0.328$m (see Fig. \ref{f1}%
c,f). Using Eq. (\ref{12}) we got for PDD 
\begin{equation}
\Delta ^{\left( 2\right) }\phi =0.530535\text{rad,}  \label{14}
\end{equation}%
which is less than the value (\ref{2.7}) obtained in the article \cite{c3,c4}%
, by 3.2\%. The difference seems to be related to the fact that in these
calculations, the contributions from platforms and other sources of gravity
were not taken into account. Table S.I \cite{c16.1} contains relative
contributions to the PDD from two configurations, besides the phase values,
linear and quadratic terms in the relative phase variations, due to the
uncertainties of atomic coordinates and velocities, obtained using\ Eqs. (%
\ref{7}, \ref{7.1}, \ref{13}), are also given. We used the value of the $zz-$%
component of the gravity gradient tensor of the Earth field, $\Gamma
_{E33}=3.11\cdot 10^{-6}$s$^{-2},$ measured in the article \cite{c1.6}.
Using data from the Table S.I and Eqs. (\ref{6a}, \ref{6b}, \ref{10}, \ref%
{10.1}), we obtained Eqs. (\ref{S.1a}, \ref{S.1b}) for the RSD and shift.
For SDs achieved in \cite{c3,c4} 
\begin{subequations}
\label{17}
\begin{eqnarray}
\sigma \left( x_{jI}\right) &=&\sigma \left( y_{jI}\right) =10^{-3}\text{m},
\label{17a} \\
\sigma \left( z_{jI}\right) &=&10^{-4}\text{m},  \label{17b} \\
\sigma \left( v_{xjI}\right) &=&\sigma \left( v_{yjI}\right) =6\cdot 10^{-3}%
\text{m/s},  \label{17c} \\
\sigma \left( v_{zjI}\right) &=&3\cdot 10^{-3}\text{m/s,}  \label{17d}
\end{eqnarray}%
one arrives to the RSD and the shift 
\end{subequations}
\begin{subequations}
\label{18}
\begin{eqnarray}
\sigma \left( \delta \Delta _{s}^{\left( 2\right) }\phi \right) &=&275\text{%
ppm}\left[ 1+6.14\cdot 10^{13}\left( \Gamma _{E31}^{2}+\Gamma
_{E32}^{2}\right) \right] ^{1/2}\text{,}  \label{18a} \\
s\left( \delta \Delta _{s}^{\left( 2\right) }\phi \right) &=&199\text{ppm.}
\label{18b}
\end{eqnarray}%
The non-diagonal matrix elements of the gradient tensor of the Earth's field
consist of three contributions arising from the fact that the Geoid is not
spherical, from the rotation of the Earth, and from the anomalous part of
the field. The first two contributions were taken into account exactly \cite%
{c1.7}, and they are $3$ orders of magnitude smaller than the diagonal
element $\Gamma _{E33}$. We have not been able to find any information about
the anomalous part of the Earth's gravitational field. However, it is seen
that the non-diagonal elements of the tensor can be neglected with an
accuracy of not more than 10\% if 
\end{subequations}
\begin{equation}
\sqrt{\Gamma _{E31}^{2}+\Gamma _{E32}^{2}}<58.5E.  \label{18.1}
\end{equation}

\section{Phase double difference and error budget at optimal conditions}

Let us apply now for the system of 24 cylinders the optimization procedure
proposed in \cite{c5,c10}. It should be noted that this procedure can be
implemented only if the technique of eliminating the gradient of the Earth's
gravitational field \cite{c17} has been previously applied. According to 
\cite{c12} and unlike \cite{c3,c4} we have chosen for calculations in this
section the distance between the lower and upper set of cylinders $dh=0.05$%
m. Our first task is to find the points of maximum and minimum (in the space
of coordinates and velocities $\left\{ z,v_{z}\right\} $) of the phase
produced by the source mass field in $C-$configuration, $\phi _{s}^{\left(
C\right) }.$ Putting $T_{1}=0,$ we found\ for these points (see Fig. \ref{f2}%
) 
\begin{subequations}
\label{19}
\begin{eqnarray}
\left\{ z_{\max },v_{\max },\phi _{s}^{\left( C\right) }\left( z_{\max
},v_{z\max }\right) \right\} &=&\left\{ -0.124\text{m},1.563\text{m/s}%
,0.215086\right\} ,  \label{19a} \\
\left\{ z_{\min },v_{\min },\phi _{s}^{\left( C\right) }\left( z_{\min
},v_{z\min }\right) \right\} &=&\left\{ 0.261\text{m},1.563\text{m/s}%
,-0.212213\right\} .  \label{19b}
\end{eqnarray}%

\begin{figure}[!t]
\includegraphics[width=15cm]{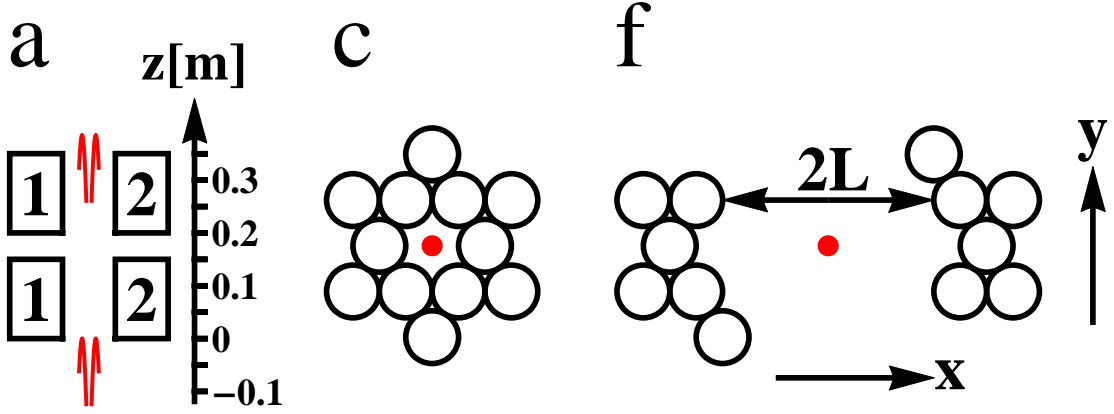}
\caption{The same as Fig. \protect\ref%
{f1}, but with different assignments for the halves 1 and 2. (a) The
cross-section $y=0$ for the $C-$ configuration, (c) and (f) top views for $%
C- $ and $F-$configurations.}
\label{f2}
\end{figure}

The phase difference of the first order will be equal to 
\end{subequations}
\begin{equation}
\Delta _{s}\phi ^{\left( C\right) }=0.427299\text{rad.}  \label{20}
\end{equation}

At this stage, before performing the calculations in the $F-$configuration,
we can calculate the shift and the RSD only with respect to the phase
difference (\ref{20}). Table S.II contains linear and quadratic terms in the
relative variation of the phase $\Delta _{s}^{\left( C\right) }$, due to the
uncertainties of atomic coordinates and velocities, obtained using\ Eqs. (%
\ref{7}, \ref{13}). One sees that despite the choice of extreme points,
linear dependences in phase variation do not completely disappear. This is
because extrema were found not exactly but approximately. If $\left\{
z,v_{z}\right\} $ is a given approximate extreme point of the function $%
f\left( z,v_{z}\right) $, then using the iteration formulas 
\begin{subequations}
\label{21}
\begin{eqnarray}
\left\{ z,v_{z}\right\} &\rightarrow &\left\{ z+\delta z,v_{z}+\delta
v_{z}\right\} ,  \label{21a} \\
\delta z &=&\left( \dfrac{\partial f}{\partial z}\dfrac{\partial ^{2}f}{%
\partial v_{z}^{2}}-\dfrac{\partial f}{\partial v_{z}}\dfrac{\partial ^{2}f}{%
\partial z^{2}}\right) \left/ \left[ \left( \dfrac{\partial ^{2}f}{\partial
z\partial v_{z}}\right) ^{2}-\dfrac{\partial ^{2}f}{\partial z^{2}}\dfrac{%
\partial ^{2}f}{\partial v_{z}^{2}}\right] \right. ,  \label{21b} \\
\delta v_{z} &=&\left( \dfrac{\partial f}{\partial v}\dfrac{\partial ^{2}f}{%
\partial z^{2}}-\dfrac{\partial f}{\partial z}\dfrac{\partial ^{2}f}{%
\partial v_{z}^{2}}\right) \left/ \left[ \left( \dfrac{\partial ^{2}f}{%
\partial z\partial v_{z}}\right) ^{2}-\dfrac{\partial ^{2}f}{\partial z^{2}}%
\dfrac{\partial ^{2}f}{\partial v_{z}^{2}}\right] \right. ,  \label{21c}
\end{eqnarray}%
one can find out the extremum with any arbitrarily high accuracy. Repeating
iterations (\ref{21}) three times we determined the velocities $v_{z\max }$
and $v_{z\min }$ with an accuracy of $10^{-7}m/s.$ These velocities match up
to the 5th digit. They are also close to the velocity of the atomic fountain 
\cite{c24} $v=gT.$ differing from it only in the third digit, 
\end{subequations}
\begin{equation}
\delta v=v_{z\max }-gT\approx -6\cdot 10^{-3}\text{m/s}  \label{21.1}
\end{equation}

This difference, however, is sufficient to exclude the parasitic signal\ 
\cite{c25}, which occurs when atoms interact with a Raman pulse having an
opposite sign of the effective wave vector. Indeed, the Raman frequency
detuning for the parasitic signal $\delta =2k\delta v\approx -2\cdot 10^{5}$s%
$^{-1}.$ If the duration of the $\pi -$pulse $\tau \thicksim 60\mu $s, then
the absolute value of the detuning $\delta $ is an order of magnitude
greater than the inverse pulse duration, and the probability of excitation
of atoms by a parasitic Raman field is negligible, is estimated to be about
4\%.

In addition, the velocity (\ref{21.1}) is twice as large as the thermal
velocity in the atomic cloud, $\bar{v}=3\cdot 10^{-3}$m$/$s \cite{c3,c4},
and therefore the portion of atoms having the opposite velocity $-\delta v$
and being in resonance with the parasitic Raman pulse is also exponentially
small, is no more than $10^{-8}$.

\begin{figure}[!t]
\includegraphics[width=15cm]{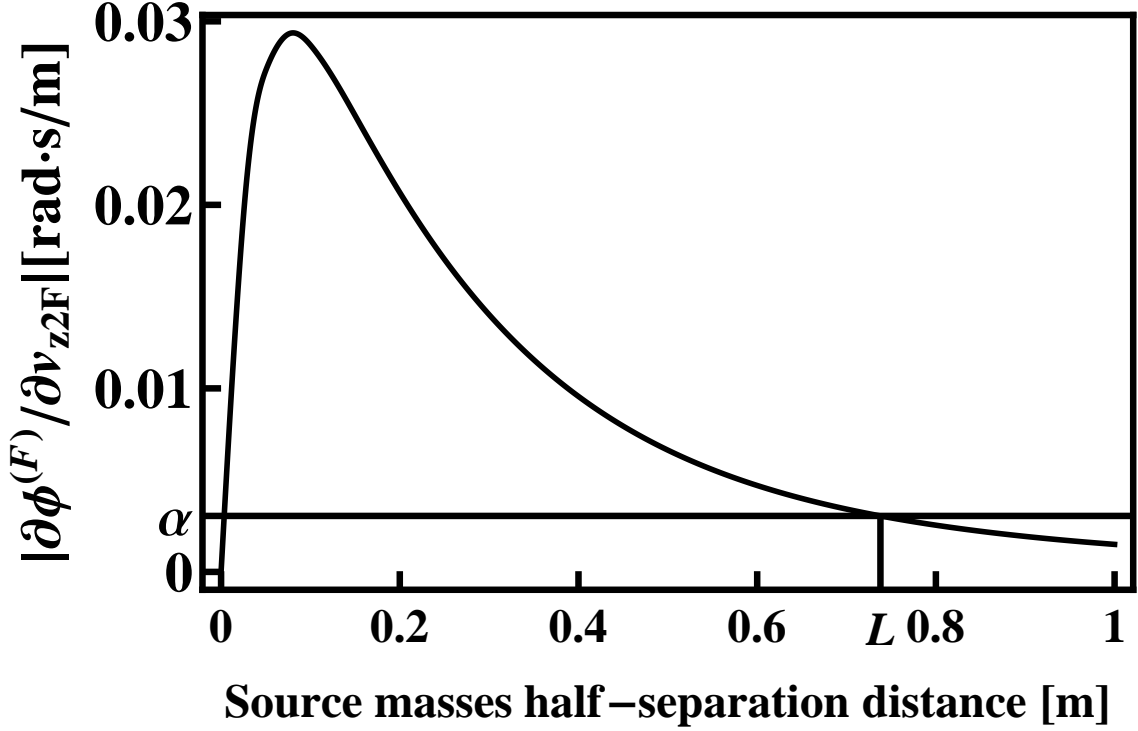}
\caption{To determine the distance $L$ by which one should move
the halves of the source mass in the $F-$configuration.}
\label{f3}
\end{figure}

Using the data from Table S.II and Eqs. (\ref{6a}, \ref{6b}, \ref{10}, \ref%
{10.1}) we got Eqs. (\ref{S.2a}, \ref{S.2b}) for RSD and relative shift. At
the uncertainties (\ref{17}) RSD and shift are given by 
\begin{subequations}
\label{24}
\begin{eqnarray}
\sigma \left( \Delta _{s}^{\left( C\right) }\phi \right) &=&93\text{ppm,}
\label{24a} \\
s\left( \Delta _{s}^{\left( C\right) }\phi \right) &=&116\text{ppm.}
\label{24b}
\end{eqnarray}

Now consider the $F-$configuration. In this configuration, points $\left\{
z_{1F},v_{z1F}\right\} =\left\{ z_{\max },v_{\max }\right\} $ and $\left\{
z_{2F},v_{z2F}\right\} =\left\{ z_{\min },v_{\min }\right\} $ found above
are not extreme, and therefore the main contribution to the variation of the
phase difference of the first order, $\delta \Delta _{s}^{\left( F\right)
}\phi $, arises from the linear terms. Including only these terms, one
receives for RSD from Eqs. (\ref{6b}, \ref{10b}) 
\end{subequations}
\begin{eqnarray}
\sigma \left( \Delta _{s}^{\left( F\right) }\phi \right) &\approx &\left\{ 
\left[ \dfrac{\partial \phi ^{\left( F\right) }\left( z_{1F},v_{z1F}\right) 
}{\partial z_{1F}}\dfrac{\sigma \left( z_{1F}\right) }{\Delta _{s}\phi
^{\left( C\right) }}\right] ^{2}+\left[ \dfrac{\partial \phi ^{\left(
F\right) }\left( z_{1F},v_{z1F}\right) }{\partial v_{z1F}}\dfrac{\sigma
\left( v_{z1F}\right) }{\Delta _{s}\phi ^{\left( C\right) }}\right]
^{2}\right.  \notag \\
&&\left. +\left[ \dfrac{\partial \phi ^{\left( F\right) }\left(
z_{2F},v_{z2F}\right) }{\partial z_{2F}}\dfrac{\sigma \left( z_{2F}\right) }{%
\Delta _{s}\phi ^{\left( C\right) }}\right] ^{2}\sigma ^{2}\left(
z_{2F}\right) +\left[ \dfrac{\partial \phi ^{\left( F\right) }\left(
z_{2F},v_{z2F}\right) }{\partial v_{z2F}}\dfrac{\sigma \left( v_{z2F}\right) 
}{\Delta _{s}\phi ^{\left( C\right) }}\right] ^{2}\right\} ^{1/2}  \label{25}
\end{eqnarray}%
The distance $2L$ between the halves of the source mass should be chosen so
large that an increase in RSD (\ref{24a}) would be insignificant.
Specifically, we demanded a full RSD did not exceed (\ref{24a}) on the
amount~more than 10\%, which means that%
\begin{equation}
\sigma \left( \Delta _{s}^{\left( F\right) }\phi \right) \leq \sqrt{0.21}%
\sigma \left( \Delta _{s}^{\left( C\right) }\phi \right) .  \label{26}
\end{equation}%
One can guarantee the fulfillment of this condition if the absolute value of
the each of four terms in square brackets in (\ref{25}) is less than $\left( 
\sqrt{0.21}/2\right) \sigma \left( \Delta _{s}^{\left( C\right) }\phi
\right) $. The most dangerous here is the last term in (\ref{25}). From the
Fig. \ref{f3} one can be sure that the inequality

\begin{equation}
\left\vert \dfrac{\partial \phi ^{\left( F\right) }\left(
z_{2F},v_{z2F}\right) }{\partial v_{z2F}}\right\vert \leq \alpha ,
\label{27}
\end{equation}%
where 
\begin{equation}
\alpha =\dfrac{\sqrt{0.21}\Delta _{s}\phi ^{\left( C\right) }}{2\sigma
\left( v_{z2F}\right) }\sigma \left( \Delta _{s}^{\left( C\right) }\phi
\right) =3.05\cdot 10^{-3},  \label{28}
\end{equation}%
is satisfied starting from the value%
\begin{equation}
L=0.738\text{m.}  \label{29}
\end{equation}%
After the halves of the source mass are moved apart by distances of $\pm L,$
we get for the phase difference 
\begin{equation}
\Delta _{s}^{\left( F\right) }\phi =8.49006\text{mrad,}  \label{30}
\end{equation}%
and therefore the PDD decreases to the value%
\begin{equation}
\Delta ^{\left( 2\right) }\phi =0.418809.  \label{31}
\end{equation}%
Calculating [with the use of Eqs. (\ref{13})] derivatives of the
interferometers' phases, one arrives to the relative contributions to the
PDD presented in the Table S.III. With the data from this table one arrives
to the final expressions (\ref{S.3a}, \ref{S.3b}) for the RSD and shift. For
the uncertainties (\ref{17}) in the atomic positions and velocities, which
we choose here following achievements in \cite{c3,c4}, one gets

\begin{subequations}
\label{33}
\begin{eqnarray}
\sigma \left( \Delta _{s}^{\left( 2\right) }\phi \right) &=&100\text{ppm,}
\label{33a} \\
s\left( \Delta _{s}^{\left( 2\right) }\phi \right) &=&118\text{ppm.}
\label{33b}
\end{eqnarray}

In this section, we considered the movements of the halves of the source
mass in the horizontal plane from the $C-$configuration to the $F-$%
configuration. Another option was considered in the article \cite{c15}, in
which it was assumed that the source mass as a whole moves in the vertical
direction. To determine the distance $L$ of the displacement in the vertical
direction, it is necessary to return to the Eq. (\ref{25}) and inequality (%
\ref{26}). The calculation showed that in this case the greatest danger is
again the last term in braces in the equation (\ref{25}). So we return to
inequality (\ref{27}). Solving it numerically, one obtains that the
displacement must be not less than 
\end{subequations}
\begin{equation}
L=1.33\text{m.}  \label{34}
\end{equation}%
Since this distance is 1.8 times greater than the distance (\ref{29}), one
concludes that horizontal displacement is preferable than vertical
displacement. I would also like to emphasize that the horizontal
displacement of the source mass was implemented in the article \cite{c9}.

\section{\label{s6}13 tons source mass}

We have already mentioned above that G. Rosi proposed and studied \cite{c15}
a new approach to the measurement of $G$ with an accuracy of 10 ppm, based
on the technique of eliminating the gravity-gradient terms \cite{c17}. In
addition to the new technique, estimates have been performed for the source
mass weight increased to the 13 tons, time separation between Raman pulses
increased to 
\begin{equation}
T=243\text{ms,}  \label{34.1}
\end{equation}%
and the uncertainty of the velocity of atomic clouds reduced to 
\begin{subequations}
\label{35}
\begin{eqnarray}
\sigma \left( v_{xjI}\right) &=&\sigma \left( v_{yjI}\right) =2\text{mm/s},
\label{35a} \\
\sigma \left( v_{zjI}\right) &=&0.3\text{mm/s.}  \label{35b}
\end{eqnarray}%
In this section we show that the optimization of PDD applied to set of
cylinders used in \cite{c3,c4}, with a total weight of about 13 tons and
uncertainties of the atomic initial positions and velocities (\ref{17a}, \ref%
{17b}) and (\ref{35}), can also lead to accuracy of PDD measurement of the
order of 10 ppm. Specifically, we assume that the cylinders are located on a
5-storey structure with a distance between the floors of $0.20011$m.
Distance between floors includes the height of the cylinder $h=0.15011$m 
\cite{c3,c4} and distance $dh=5$cm between cylinders tops and bottoms. Each
of the floors has 126 cylinders as shown in Fig. \ref{f4}c,f. This
arrangement of the cylinders is a natural generalization of the geometry
chosen in the \cite{c3,c4}.

\begin{figure}[!t]
\includegraphics[width=15cm]{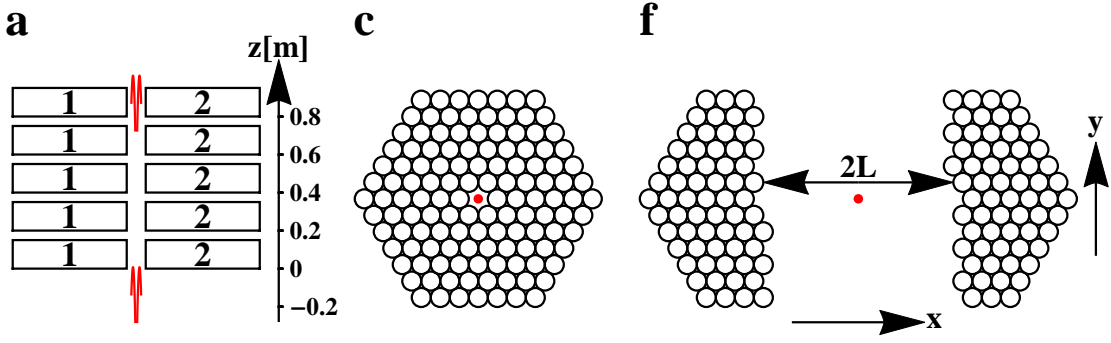}
\caption{Same
as Fig. \protect\ref{f2}, but for 630 cylinders.}
\label{f4}
\end{figure}

Maximum and minimum points of the function $\phi ^{\left( C\right) }\left(
z,v\right) ,$%
\end{subequations}
\begin{subequations}
\label{36}
\begin{eqnarray}
\left\{ z_{\max },v_{z\max },\phi _{s}^{\left( C\right) }\left( z_{\max
},v_{z\max }\right) \right\} &=&\left\{ -0.281\text{m},2.38\text{m/s},1.84859%
\text{rad}\right\} ,  \label{36a} \\
\left\{ z_{\min },v_{z\min },\phi _{s}^{\left( C\right) }\left( z_{\min
},v_{z\min }\right) \right\} &=&\left\{ 0.727\text{m},2.38\text{m/s},-1.81252%
\text{rad}\right\} ,  \label{36b}
\end{eqnarray}%
one selects for the initial coordinates and velocities of the atomic clouds
in the first and second interferometer. The phase difference of the first
order in this case is 
\end{subequations}
\begin{equation}
\Delta _{s}^{\left( C\right) }\phi =3.66111\text{rad.}  \label{37}
\end{equation}%
Linear and quadratic terms in relative variations $\delta \Delta
_{s}^{\left( C\right) }\phi /\Delta _{s}^{\left( C\right) }\phi $ are pieced
together in the Table S.IV. \ Substituting these data into the equations (%
\ref{10}), one obtains for RSD and shift Eqs. (\ref{S.4a}, \ref{S.4b}). For
uncertainties (\ref{17a}, \ref{17b}, \ref{35}) RSD and shift are equal to 
\begin{subequations}
\label{39}
\begin{eqnarray}
\sigma \left( \Delta _{s}^{\left( C\right) }\phi \right) &=&14\text{ppm,}
\label{39a} \\
s\left( \Delta _{s}^{\left( C\right) }\phi \right) &=&20\text{ppm.}
\label{39b}
\end{eqnarray}%
Let us now consider the contribution from the $F-$configuration of the
source mass. One determines the distance $L$ of the halves of the source
mass displacement in the horizontal direction from the inequality (\ref{26}%
). The calculation showed that the greatest danger comes from the third term
in braces in Eq. (\ref{25}). Then one guarantees the inequality (\ref{26})
fulfilment if 
\end{subequations}
\begin{equation}
\left\vert \dfrac{\partial \phi ^{\left( F\right) }\left(
z_{2F},v_{z2F}\right) }{\partial z_{2F}}\right\vert \leq \alpha ,  \label{40}
\end{equation}%
where%
\begin{equation}
\alpha =\dfrac{\sqrt{0.21}\Delta _{s}\phi ^{\left( C\right) }}{2\sigma
\left( z_{2F}\right) }\sigma \left( \Delta _{s}^{\left( C\right) }\phi
\right) =0.1195,  \label{41}
\end{equation}%
From Fig. \ref{f5} one sees that the halves of the source mass should be
moved apart by a distance.%
\begin{equation}
L=1.39\text{m.}  \label{42}
\end{equation}%

\begin{figure}[!t]
\includegraphics[width=15cm]{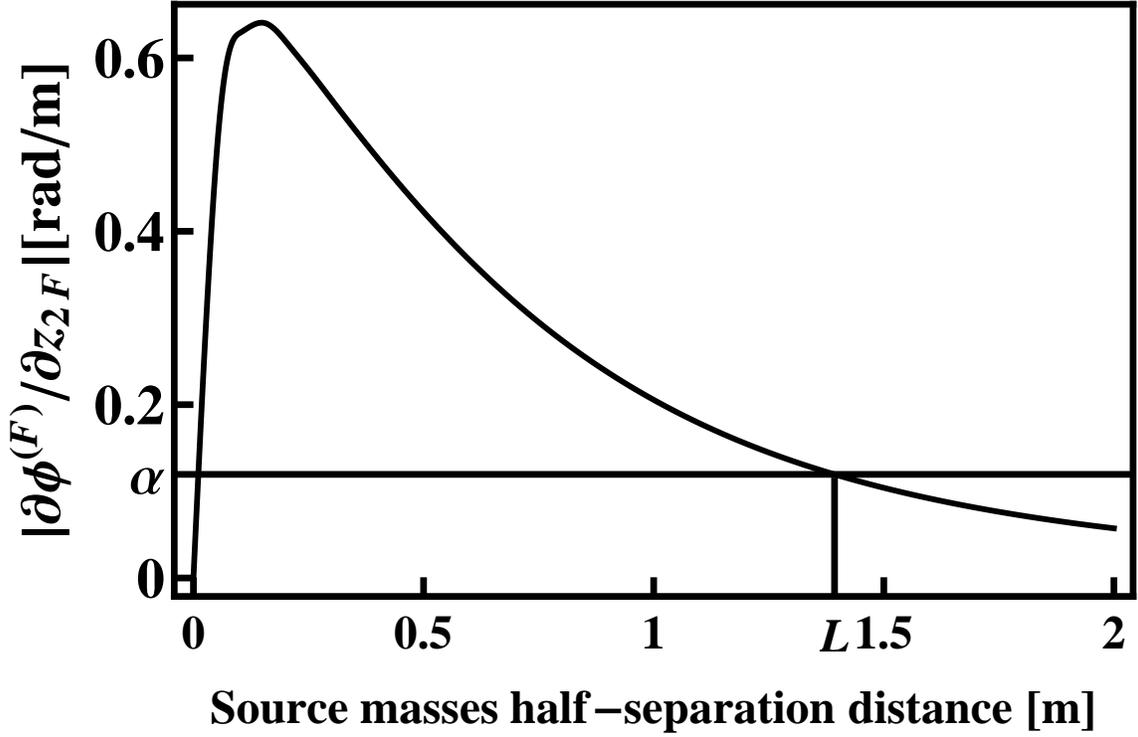}
\caption{Same as Fig. \protect\ref{f3},
but for 630 cylinders.}
\label{f5}
\end{figure}

Then, by
computing the first order phase difference in the $F-$configuration, $\Delta
_{s}\phi ^{\left( F\right) }$, one obtains for the PDD%
\begin{equation}
\Delta ^{\left( 2\right) }\phi =3.51021\text{rad}  \label{43}
\end{equation}%
With this signal, the various relative contributions to the PDD are
presented in the table S.V. Calculating from these data the first and second
order derivatives with respect to the coordinates and velocities of the
atomic clouds, and substituting these derivatives in the equations (\ref{10}%
), one arrives at the Eqs. (\ref{S.5a}, \ref{S.5b}), from which at the
uncertainties (\ref{17a}, \ref{17b}, \ref{35}), RSD and shift are equal to 
\begin{subequations}
\label{45}
\begin{eqnarray}
\sigma \left( \Delta _{s}^{\left( 2\right) }\phi \right) &=&16\text{ppm,}
\label{45a} \\
s\left( \Delta _{s}^{\left( 2\right) }\phi \right) &=&20\text{ppm.}
\label{45b}
\end{eqnarray}

Note also that we have considered the displacement of the source mass as a
whole in the vertical direction, and found that the minimal distance at this
displacement is equal to 
\end{subequations}
\begin{equation}
L=2.92\text{m.}  \label{46}
\end{equation}%
This displacement is only 25\% shorter than the displacement of the source
mass by 4m required for the technique \cite{c15}. However, it is more than
two times longer than the horizontal displacement (\ref{42}). Therefore, in
the case of a 13-ton source mass, horizontal displacement is preferable.

\section{\label{s7}Conclusion}

This article is devoted to the calculation of the error budget in the
measurement of the Newtonian gravitational constant $G$ by atomic
interferometry methods. Using the technique \cite{c14}, we obtained
expressions for the gravitational field of the cylinder, which is used in
these measurements.

Despite the compensation of the gradient of the Earth gravitational field at
the points of apogees of the atomic trajectories achieved in the article 
\cite{c3,c4}, an absence of this compensation along the entire trajectory
leads to the influence of the Earth's field on the $G$ measurement accuracy.
To overcome this influence, we propose to use the method of eliminating the
gradient of the gravitational field of the Earth \cite{c17}.

The main attention in this article is paid to the calculation of standard
deviation (SD) and the shift of the PDD due to the uncertainties of the mean
values of the initial coordinates and the velocities of atomic clouds $%
\left\{ \delta \mathbf{x},\delta \mathbf{v}\right\} $. We propose to include
in the error budget new terms. They are originated from the quadratic
dependence of the variation of the AI phases on $\left\{ \delta \mathbf{x}%
,\delta \mathbf{v}\right\} $. The shift arises only after including those
terms. At the conditions realized in the article \cite{c3,c4}, calculations
brings us to the shift (\ref{18b}) and to the opposite relative correction $%
\Delta G/G=-199$ppm, which is larger than corrections considered in \cite%
{c3,c4}. After including this correction, the value of the gravitational
constant $G$ should be shifted to 
\begin{equation}
G=6.67058\text{m}^{3}\text{kg}^{-1}\text{s}^{-2}  \label{46.1}
\end{equation}%
from the value $G=6.67191$m$^{3}$kg$^{-1}$s$^{-2}$ measured in \cite{c3,c4}.
We would like to note that Monte Carlo simulation in \cite{c3,c4} did not
bring to any shift caused by uncertainties of the atom position and velocity.

After eliminating the gradient of the Earth field, one can use optimization
technique proposed in \cite{c5,c10}. While the PDD in this case becomes
smaller [compare signals (\ref{14}) and (\ref{20})], one should be able to
measure $G$ with $2.75$ times better accuracy [compare RSDs (\ref{18a}) and (%
\ref{33a})].

Another application of our formulas is the calculation of the systematic
error due to the finite size of atomic clouds and their finite temperature 
\cite{c5,c10}. Let us now assume that $\delta \mathbf{x}$ is the deviation
of the atom from the center of the cloud and $\delta \mathbf{v}$ is the
deviation from the center of the atomic velocity distribution. If the
temperatures are small enough to ignore the Doppler frequency shift, and the
aperture of the optical field is large enough to assume that the areas of
the Raman pulses do not depend on the position of atoms in the cloud, then
the only reason for the dependence of the PDD on $\left\{ \delta \mathbf{x}%
,\delta \mathbf{v}\right\} $ is that the gravitational field $\delta \mathbf{%
g}\left[ \mathbf{x}\left( t\right) \right] $ is not the same for different
atoms in the cloud. Averaging the phase of the AI over an atomic
distribution one will receive from the equation (\ref{7}) 
\begin{equation}
\left\langle \delta \phi \right\rangle =\dfrac{1}{4}\left( a_{m}^{2}\dfrac{%
\partial ^{2}\phi }{\partial x_{m}^{2}}+v_{0m}^{2}\dfrac{\partial ^{2}\phi }{%
\partial v_{m}^{2}}\right) ,  \label{47}
\end{equation}%
where $a_{m}=\sqrt{2\left\langle \delta x_{m}^{2}\right\rangle }$ and $%
v_{0m}=\sqrt{2\left\langle \delta v_{m}^{2}\right\rangle }$ are the radius
and thermal velocity of the atomic cloud along the $m-$axis. Here we pay
attention to the fact that at equal radii, $a_{x}=a_{y}=a_{z},$ and
temperatures, $v_{0x}=v_{0y}=v_{0z},$ the systematic error (\ref{47})
disappears. This follows from the Eqs. (\ref{13c}, \ref{13e}) and from the
fact that the gravitational field obeys the Laplace equation and, therefore,
the trace of the gravitational field curvature tensor (\ref{13.1}) $\chi
_{s3mm}=0$.

In the absence of linear terms, one can diagonalize the quadratic shape in
phase variation (\ref{7}), so that.%
\begin{equation}
\delta \phi =\dsum_{m=1}^{6}a_{m}\delta q_{m}^{2}.  \label{48}
\end{equation}%
By analogy with the error budget in \cite{c2,c3,c4,c1.4,c12,c15,c16}, if
each term $m$ in Eq. (\ref{48}) is independent of the others, one might
expect that his contribution to the SD is 
\begin{equation}
\sigma _{m}\left( \phi \right) =\left\vert a_{m}\right\vert \sigma
^{2}\left( q_{m}\right) ,  \label{48.1}
\end{equation}%
and%
\begin{equation}
\sigma \left( \phi \right) =\left[ \dsum_{m=1}^{6}\sigma _{m}^{2}\left( \phi
\right) \right] ^{1/2}.  \label{49}
\end{equation}%
This, however, is not precisely correct. The calculation shows that for
independent extreme variables, the expression (\ref{49}) is true, but the
contribution of each term is%
\begin{equation}
\sigma _{m}\left( \phi \right) =\sqrt{2}\left\vert a_{m}\right\vert \sigma
^{2}\left( q_{m}\right) \left( 1+\dfrac{1}{2}\kappa \left( q_{m}\right)
\right) ^{1/2},  \label{50}
\end{equation}%
where $\kappa \left( q_{m}\right) $ is a cumulant (\ref{9}), i.e. even for
normal distribution, when $\kappa \left( q_{m}\right) =0$, the contribution
from each term is $\sqrt{2}$ times greater than in (\ref{48.1}).

Instead of using tables, we propose to use for error budget expressions for
RSD (\ref{S.1a} - \ref{S.5a}) and relative shift (\ref{S.1b} - \ref{S.5b})
[obtained by means of general Eqs. (\ref{10})], in which one should
substitute the coordinates and velocities uncertainties achieved or assumed
to be achieved in the experiment.

In this article we have only considered error budget related to atomic
variables and did not consider errors associated with the properties of the
source mass. Examples of calculation errors associated with the
uncertainties of positioning and orientation of the various components of
the source mass can be found in the article \cite{c10}. We would only like
to emphasize that, as in the article \cite{c15}, it is preferable to use
source masses consisting of as many components as possible. So if one uses N
cylinders, the signal $f$ consists of the contributions $f_{m}$ of the each
cylinder $\left( m=1,\ldots N\right) .$ For RSD one gets $\sigma _{r}\left(
f\right) =\left. \left[ \dsum_{m}f_{m}^{2}\sigma _{r}^{2}\left( f_{m}\right) %
\right] ^{1/2}\right/ f$. If one accepts for estimates that the
contributions of the various cylinders and their RSDs are of the same order, 
$f_{m}\thicksim f_{m^{\prime }}\thicksim f/N$ and $\sigma _{r}\left(
f_{m}\right) \thicksim \sigma _{r}\left( f_{m^{\prime }}\right) $, then we
get that the RSD of the entire system of cylinders decreases as $N^{-1/2}$, 
\begin{equation}
\sigma _{r}\left( f\right) \thicksim N^{-1/2}\sigma _{r}\left( f_{m}\right) .
\label{52}
\end{equation}

Finally, we would like to note that, following the statement \cite{c15},
that 13-ton source mass can be implemented in the experiment, we increased
the number of cylinders to $630$ (more than $26$ times). At the same time,
the optimal signal increased only $8.4$ times [compare Eqs. (\ref{43}) and (%
\ref{31})], and this increase is partly due to an increase in the delay time
between the Raman pulses $T.$ This example shows that an increase in the
weight of the source mass does not even lead to a proportional signal
increase. More promising here is an increase in the signal due to the large
values of $T$, the effective wave vector $k$ and the optimal aspect ratio of
the source mass. Due to these factors we predicted \cite{c5} PDD $\Delta
_{s}^{\left( 2\right) }\phi =386.527$rad even for a source mass $M=1080$kg.

\acknowledgments Author is appreciated to Dr. M. Prevedeli for the fruitful
discussions and to Dr. G. Rosi for the explanation of some points in the
article \cite{c15}.

\appendix

\section{\label{sa}Gravity field of the homogeneous cylinder}

\subsection{Axial component}

It is convenient \cite{c14} to explore the following expression for the
potential of the gravitational field of a homogeneous cylinder $\Phi \left( 
\mathbf{x}\right) $ 
\begin{equation}
\Phi \left( r,z\right) =-2G\rho \int_{0}^{R}dy\int_{r-\sqrt{R^{2}-y^{2}}}^{r+%
\sqrt{R^{2}-y^{2}}}d\xi \int_{z-h}^{z}\dfrac{d\zeta }{\sqrt{y^{2}+\xi
^{2}+\zeta ^{2}}},  \label{a1}
\end{equation}%
where $\rho ,$ $R,$ and $h$ are the density, radius, and height of the
cylinder, $\left( r,z,\psi =0\right) $ are the cylindrical coordinates of
the vector $\mathbf{x}$.. For an axial component of the gravitational field, 
$\delta g_{3}\left( r,z\right) =-\partial _{z}\Phi \left( r,z\right) $, one
gets%
\begin{equation}
\delta g_{3}\left( r,z\right) =2G\rho g_{3}\left( r,\zeta \right) _{\zeta
=z-h}^{\zeta =z},  \label{a2}
\end{equation}%
where the function 
\begin{equation}
g_{3}\left( r,\zeta \right) =\int_{0}^{R}dy\int_{r-\sqrt{R^{2}-y^{2}}}^{r+%
\sqrt{R^{2}-y^{2}}}\dfrac{d\xi }{\sqrt{y^{2}+\xi ^{2}+\zeta ^{2}}}
\label{a3}
\end{equation}%
can be represented as 
\begin{subequations}
\label{a4}
\begin{eqnarray}
g_{3}\left( r,\zeta \right) &=&\int_{0}^{R}dy\ln \dfrac{t_{+}\left( y\right) 
}{t_{-}\left( y\right) }  \notag \\
&=&-\int_{0}^{R}y\left( \dfrac{dt_{+}}{t_{+}}-\dfrac{dt_{-}}{t_{-}}\right) ,
\label{a4b} \\
t_{\pm }\left( y\right) &=&r\pm \sqrt{R^{2}-y^{2}}+\left( \zeta
^{2}+r^{2}+R^{2}\pm 2r\sqrt{R^{2}-y^{2}}\right) ^{1/2}  \label{a4c}
\end{eqnarray}%
Since $t_{+}\left( R\right) =t_{-}\left( R\right) \equiv t\left( R\right)
\lessgtr t_{\pm }\left( 0\right) $ one can write 
\end{subequations}
\begin{equation}
g_{3}\left( r,\zeta \right) =\int_{t\left( R\right) }^{t_{+}\left( 0\right) }%
\dfrac{dt}{t}y_{+}\left( t\right) +\int_{t_{-}\left( 0\right) }^{t\left(
R\right) }\dfrac{dt}{t}y_{-}\left( t\right) ,  \label{a5}
\end{equation}%
where $y_{\pm }\left( t\right) $ is the root of the equation $t_{\pm }\left(
y\right) =t$. To find this root, consider the functions $x_{\pm }\left(
t\right) =\sqrt{R^{2}-y_{\pm }^{2}\left( t\right) },$.%
\begin{equation}
0<x_{\pm }\left( t\right) <R.  \label{a6}
\end{equation}%
For them one gets%
\begin{equation}
x_{\pm }\left( t\right) =\pm t+\sqrt{\zeta ^{2}+R^{2}+2tr}\text{ or }\pm t-%
\sqrt{\zeta ^{2}+R^{2}+2tr}  \label{a7}
\end{equation}%
Since $t+\sqrt{\zeta ^{2}+R^{2}+2tr}>R$, then one should choose $x_{+}\left(
t\right) =t-\sqrt{\zeta ^{2}+R^{2}+2tr}.$. Since $t_{-}\left( 0\right)
>r-R+\left\vert r-R\right\vert >0,-t-\sqrt{\zeta ^{2}+R^{2}+2tr}<0$, hence $%
x_{-}\left( t\right) =\sqrt{\zeta ^{2}+R^{2}+2tr}-t$ or%
\begin{equation}
x_{\pm }\left( t\right) =\pm \left( t-\sqrt{\zeta ^{2}+R^{2}+2tr}\right) .
\label{a8}
\end{equation}%
Therefore, one concludes that the functions $y_{\pm }\left( t\right) $ are
coincident and equal to%
\begin{equation}
y_{+}\left( t\right) =y_{-}\left( t\right) =y\left( t\right) =\left[
2t\left( \sqrt{\zeta ^{2}+R^{2}+2tr}-r\right) -t^{2}-\zeta ^{2}\right] ^{1/2}
\label{a9}
\end{equation}%
and.%
\begin{equation}
g_{3}\left( r,\zeta \right) =\int_{t_{-}\left( 0\right) }^{t_{+}\left(
0\right) }\dfrac{dt}{t}y\left( t\right) .  \label{a10}
\end{equation}%
Introducing new variable,%
\begin{equation}
u=\sqrt{\zeta ^{2}+R^{2}+2tr}-r,  \label{a11}
\end{equation}%
for which 
\begin{subequations}
\label{a12}
\begin{eqnarray}
u\left[ t_{\pm }\left( 0\right) \right] &\equiv &u_{\pm }=\sqrt{\zeta
^{2}+\left( r\pm R\right) ^{2}},  \label{a12a} \\
y\left( t\right) &=&\dfrac{\sqrt{q\left( u^{2}\right) }}{2r},  \label{a12b}
\\
q\left( \eta \right) &=&\left\{ u_{+}^{2}-\eta \right\} \left\{ \eta
-u_{-}^{2}\right\} ,  \label{a12c} \\
dt &=&\dfrac{u+r}{r}du  \label{a12d}
\end{eqnarray}%
and so 
\end{subequations}
\begin{subequations}
\label{a13}
\begin{eqnarray}
g_{3}\left( r,\zeta \right) &=&I+I^{\prime },  \label{a13a} \\
I &=&\int_{u_{-}}^{u_{+}}\dfrac{du}{\sqrt{q\left( u^{2}\right) }}J\left(
u\right) ,  \label{a13b} \\
J\left( u\right) &=&\dfrac{q\left( u^{2}\right) \left( r^{2}-\zeta
^{2}-R^{2}-u^{2}\right) }{w\left( u^{2}\right) },  \label{a13c} \\
I^{\prime } &=&\dfrac{1}{2r}\int_{u_{-}^{2}}^{u_{+}^{2}}d\eta J^{\prime
}\left( \eta \right) ,  \label{a13d} \\
J^{\prime }\left( \eta \right) &=&\dfrac{\left( \eta -\zeta
^{2}-R^{2}-r^{2}\right) }{\sqrt{q\left( \eta \right) }w\left( \eta \right) },
\label{a13e} \\
w\left( \eta \right) &=&\left( \eta -\eta _{1}\right) \left( \eta -\eta
_{2}\right) ,  \label{a13f} \\
\eta _{1,2} &=&\left( r\pm \sqrt{\zeta ^{2}+R^{2}}\right) ^{2}.  \label{a13g}
\end{eqnarray}%
Using equality 
\end{subequations}
\begin{equation}
w\left( \eta \right) +q\left( \eta \right) =-4r^{2}\zeta ^{2},  \label{a14}
\end{equation}%
one can show that the integrand $J^{\prime }\left( \eta \right) $ is an
antisymmetric function with respect to the middle point $\eta =\left\{ u^{2}%
\left[ t_{+}\left( 0\right) \right] +u^{2}\left[ t_{-}\left( 0\right) \right]
\right\} /2,$ and, therefore, the term (\ref{a13d}) is equal $0$. At the
same time, expanding $J\left( u\right) $ into partial fractions, one obtains 
\begin{subequations}
\label{a15}
\begin{eqnarray}
g_{3}\left( r,\zeta \right) &=&\left( R^{2}+\zeta ^{2}-r^{2}\right)
I_{1}+I_{2}+I_{3+}+I_{3-},  \label{a15a} \\
I_{1} &=&\int_{u_{-}}^{u_{+}}\dfrac{du}{\sqrt{q\left( u^{2}\right) }},
\label{a15b} \\
I_{2} &=&\int_{u_{-}}^{u_{+}}\dfrac{duu^{2}}{\sqrt{q\left( u^{2}\right) }},
\label{a15c} \\
I_{3\pm } &=&2r\zeta ^{2}\left( r\pm \sqrt{\zeta ^{2}+R^{2}}\right)
\int_{u_{-}}^{u_{+}}\dfrac{du}{\sqrt{q\left( u^{2}\right) }\left( u^{2}-\eta
_{1,2}\right) }  \label{a15d}
\end{eqnarray}%
The integrals (\ref{a15}), one can compute using the substitution 
\end{subequations}
\begin{equation}
u=\sqrt{u_{+}^{2}-\left( u_{+}^{2}-u_{-}^{2}\right) \sin ^{2}\phi }
\label{a15.1}
\end{equation}%
Finally, one arrives at the following expression for the axial component of\
the cylinder's field 
\begin{subequations}
\label{a16}
\begin{eqnarray}
\delta g_{3}\left( r,z\right) &=&2G\rho g_{3}\left( r,\zeta \right) _{\zeta
=z-h}^{\zeta =z},  \label{a16a} \\
g_{3}\left( r,\zeta \right) &=&\dfrac{\left( \zeta ^{2}+R^{2}-r^{2}\right) }{%
\sqrt{\zeta ^{2}+\left( r+R\right) ^{2}}}K\left( k\right) +\sqrt{\zeta
^{2}+\left( r+R\right) ^{2}}E\left( k\right)  \notag \\
&&+\dfrac{\zeta ^{2}}{\sqrt{\zeta ^{2}+\left( r+R\right) ^{2}}}\dsum_{j=\pm
1}\left[ \dfrac{r+j\sqrt{\zeta ^{2}+R^{2}}}{R-j\sqrt{\zeta ^{2}+R^{2}}}\Pi
\left( \dfrac{2R}{R-j\sqrt{\zeta ^{2}+R^{2}}}|k\right) \right] ,
\label{a16b} \\
k &=&\sqrt{\dfrac{4rR}{\zeta ^{2}+\left( r+R\right) ^{2}}},  \label{a16c}
\end{eqnarray}%
where $K\left( k\right) ,$ $E\left( k\right) $ and $\Pi \left( \alpha
|k\right) $ are the complete elliptic integrals of the first, second and
third order respectively.

\subsection{Radial component}

For the radial component of the gravitational field $\delta g_{r}\left(
r,z\right) =-\partial _{r}\Phi \left( r,z\right) $ one obtains from (\ref{a1}%
) 
\end{subequations}
\begin{subequations}
\label{a17}
\begin{eqnarray}
\delta g_{r}\left( r,z\right) &=&2G\rho g_{r}\left( r,\zeta \right) _{\zeta
=z-h}^{\zeta =z},  \label{a17a} \\
g_{r}\left( r,\zeta \right) &=&-\int_{0}^{R}y\left( \dfrac{dt_{+}}{t_{+}}-%
\dfrac{dt_{-}}{t_{-}}\right) ,  \label{a17b} \\
t_{\pm }\left( y\right) &=&\zeta +\left[ \zeta ^{2}+r^{2}+R^{2}\pm 2r\sqrt{%
R^{2}-y^{2}}\right] ^{1/2}  \label{a17c}
\end{eqnarray}%
Since still $t_{+}\left( R\right) =t_{-}\left( R\right) \lessgtr t_{\pm
}\left( 0\right) ,$ one gets, 
\end{subequations}
\begin{equation}
g_{r}\left( r,\zeta \right) =\int_{t_{-}\left( 0\right) }^{t\left( R\right) }%
\dfrac{dt}{t}y_{-}\left( t\right) +\int_{t\left( R\right) }^{t_{+}\left(
0\right) }\dfrac{dt}{t}y_{+}\left( t\right) ,  \label{a17.1}
\end{equation}%
where $y_{\pm }\left( t\right) $ are functions inverse to (\ref{a17c}).
Since these functions are the same%
\begin{equation}
y_{+}\left( t\right) =y_{-}\left( t\right) \equiv y\left( t\right) =\dfrac{1%
}{2r}\left[ 4r^{2}R^{2}-\left( t^{2}-2\zeta t-r^{2}-R^{2}\right) ^{2}\right]
^{1/2},  \label{a18}
\end{equation}%
then, choosing as an integration variable $u=t-\zeta $, one finds that 
\begin{subequations}
\label{a19}
\begin{eqnarray}
g_{r}\left( r,\zeta \right) &=&I+I^{\prime },  \label{a19a} \\
I &=&-\dfrac{\zeta }{2r}\int_{u_{-}}^{u_{+}}\dfrac{duq\left( u^{2}\right) }{%
\left( u^{2}-\zeta ^{2}\right) \sqrt{q\left( u^{2}\right) }},  \label{a19b}
\\
I^{\prime } &=&\dfrac{1}{4r}\int_{u_{-}^{2}}^{u_{+}^{2}}\dfrac{d\eta \sqrt{%
q\left( \eta \right) }}{\left( \eta -\zeta ^{2}\right) },  \label{a19c}
\end{eqnarray}%
where $u_{\pm }$ and $q\left( \eta \right) $ are given by Eqs. (\ref{a12a}, %
\ref{a12c})$.$ Because $u_{\pm }^{2}-\zeta ^{2}$ and $q\left( \eta +\zeta
^{2}\right) $ are independent of $\zeta $, the term $I^{\prime }$ gives no
contribution to the acceleration (\ref{a17a}) and can be omitted. While
using the substitution (\ref{a15.1}), one reduces the integral in (\ref{a19b}%
) to elliptic integrals, which brings us to the next final result

\end{subequations}
\begin{subequations}
\label{a20}
\begin{gather}
\delta g_{r}\left( r,z\right) =2G\rho g_{r}\left( r,\zeta \right) _{\zeta
=z-h}^{\zeta =z},  \label{a20a} \\
g_{r}\left( r,\zeta \right) =\dfrac{\zeta }{2r\sqrt{\zeta ^{2}+\left(
r+R\right) ^{2}}}\left[ -\left( \zeta ^{2}+2r^{2}+2R^{2}\right) K\left(
k\right) +\left( \zeta ^{2}+\left( r+R\right) ^{2}\right) E\left( k\right) +%
\dfrac{\left( r^{2}-R^{2}\right) ^{2}}{\left( r+R\right) ^{2}}\Pi \left( 
\dfrac{4rR}{\left( r+R\right) ^{2}}|k\right) \right] ,  \label{a20b}
\end{gather}%
where $k$ is given by Eq. (\ref{a16c}).

\newpage

\begin{center}
{\Large \textbf{Supplemental Material}}
\end{center}

Linear and non-linear terms in the variations of the phase differences and
expressions for the relative standard deviations (RSD) and relative shifts,
obtained numerically are pieced together in the following Tables and
expressions.

\begin{center}
TABLE S.I: Relative contributions to the phase double difference (PDD) and
error budget for two configurations of source mass

.%
\begin{tabular}{|c|c|c|}
\hline
Term & C-configuration & F-configuration \\ \hline
$\pm \Delta _{s}\phi ^{I}/\Delta ^{\left( 2\right) }\phi $ & $0.685376$ & $%
0.314624$ \\ \hline
Linear in position & $%
\begin{array}{c}
0.322\delta z_{1C}+0.117\delta z_{2C}+7.77\cdot 10^{5} \\ 
\times \left[ \Gamma _{E31}\left( \delta x_{1C}-\delta x_{2C}\right) +\Gamma
_{E32}\left( \delta y_{1C}-\delta y_{2C}\right) \right]%
\end{array}%
$ & $%
\begin{array}{c}
0.132\delta z_{1F}+0.518\delta z_{2F}-7.77\cdot 10^{5} \\ 
\times \left[ \Gamma _{E31}\left( \delta x_{1F}-\delta x_{2F}\right) +\Gamma
_{E32}\left( \delta y_{1F}-\delta y_{2F}\right) \right]%
\end{array}%
$ \\ \hline
Linear in velocity & $%
\begin{array}{c}
0.0377\delta v_{z1C}+0.0150\delta v_{z2C}+1.24\cdot 10^{5} \\ 
\times \left[ \Gamma _{E31}\left( \delta v_{x1C}-\delta v_{x2C}\right)
+\Gamma _{E32}\left( \delta v_{y1C}-\delta v_{y2C}\right) \right]%
\end{array}%
$ & $%
\begin{array}{c}
0.0132\delta v_{z1F}+0.0683\delta v_{z2F}-1.24\cdot 10^{5} \\ 
\times \left[ \Gamma _{E31}\left( \delta v_{x1F}-\delta v_{x2F}\right)
+\Gamma _{E32}\left( \delta v_{y1F}-\delta v_{y2F}\right) \right]%
\end{array}%
$ \\ \hline
Nonlinear in position & $%
\begin{array}{c}
12.3\left( \delta x_{1C}^{2}+\delta y_{1C}^{2}\right) -24.7\delta z_{1C}^{2}
\\ 
+12.2\left( \delta x_{2C}^{2}+\delta y_{2C}^{2}\right) -24.3\delta z_{2C}^{2}%
\end{array}%
$ & $%
\begin{array}{c}
15.5\left( \delta x_{1F}^{2}+\delta y_{1F}^{2}\right) -30.9\delta z_{1F}^{2}
\\ 
+15.7\left( \delta x_{2F}^{2}+\delta y_{2F}^{2}\right) -31.3\delta z_{2F}^{2}%
\end{array}%
$ \\ \hline
Nonlinear in velocity & $%
\begin{array}{c}
0.375\left( \delta v_{x1C}^{2}+\delta v_{y1C}^{2}\right) -0.750\delta
v_{z1C}^{2} \\ 
+0.351\left( \delta v_{x2C}^{2}+\delta v_{y2C}^{2}\right) -0.702\delta
v_{z2C}^{2}%
\end{array}%
$ & $%
\begin{array}{c}
0.451\left( \delta v_{x1F}^{2}+\delta v_{y1F}^{2}\right) -0.901\delta
v_{z1F}^{2} \\ 
+0.468\left( \delta v_{x2F}^{2}+\delta v_{y2F}^{2}\right) -0.937\delta
v_{z2F}^{2}%
\end{array}%
$ \\ \hline
$%
\begin{tabular}{c}
Position-velocity \\ 
cross term%
\end{tabular}%
$ & $%
\begin{array}{c}
3.99\left( \delta v_{x1C}\delta x_{1C}+\delta v_{y1C}\delta y_{1C}\right)
-7.97\delta v_{z1C}\delta z_{1C} \\ 
+4.05\left( \delta v_{x2C}\delta x_{2C}+\delta v_{y2C}\delta y_{2C}\right)
-8.10\delta v_{z2C}\delta z_{2C}%
\end{array}%
$ & $%
\begin{array}{c}
5.10\left( \delta v_{x1F}\delta x_{1F}+\delta v_{y1F}\delta y_{1F}\right)
-10.2\delta v_{z1F}\delta z_{1F} \\ 
+5.08\left( \delta v_{x2F}\delta x_{2F}+\delta v_{y2F}\delta y_{2F}\right)
-10.2\delta v_{z2F}\delta z_{2F}%
\end{array}%
$ \\ \hline
\end{tabular}
\end{center}

\end{subequations}
\begin{subequations}
\label{S.1}
\begin{align}
\sigma \left( \Delta _{s}^{\left( 2\right) }\phi \right) & =\left\{
0.104\sigma ^{2}\left( z_{1C}\right) +0.0137\sigma ^{2}\left( z_{2C}\right)
+1.42\cdot 10^{-3}\sigma ^{2}\left( v_{z}{}_{1C}\right) +2.24\cdot
10^{-4}\sigma ^{2}\left( v_{z2C}\right) \right.  \notag \\
& +0.0173\sigma ^{2}\left( z_{1F}\right) +0.269\sigma ^{2}\left(
z_{2F}\right) +1.75\cdot 10^{-4}\sigma ^{2}\left( v_{z1F}\right) +4.66\cdot
10^{-3}\sigma ^{2}\left( v_{z2F}\right)  \notag \\
& +\dsum_{j=1,2}\dsum_{I=C,F}\left[ 6.04\cdot 10^{11}\left( \Gamma
_{E31}^{2}\sigma ^{2}\left( x_{jI}\right) +\Gamma _{E32}^{2}\sigma
^{2}\left( y_{jI}\right) \right) +1.55\cdot 10^{10}\left( \Gamma
_{E31}^{2}\sigma ^{2}\left( v_{xjI}\right) +\Gamma _{E32}^{2}\sigma
^{2}\left( v_{yjI}\right) \right) \right]  \notag \\
& +305\left[ \sigma ^{4}\left( x_{1C}\right) +\sigma ^{4}\left(
y_{1C}\right) \right] +1220\sigma ^{4}\left( z_{1C}\right) +296\left[ \sigma
^{4}\left( x_{2C}\right) +\sigma ^{4}\left( y_{2C}\right) \right]
+1180\sigma ^{4}\left( z_{2C}\right)  \notag \\
& +0.282\left[ \sigma ^{4}\left( v_{x1C}\right) +\sigma ^{4}\left(
v_{y1C}\right) \right] +1.13\sigma ^{4}\left( v_{z1C}\right) +0.247\left[
\sigma ^{4}\left( v_{x2C}\right) +\sigma ^{4}\left( v_{y2C}\right) \right]
+0.987\sigma ^{4}\left( v_{z2C}\right)  \notag \\
& +15.9\left[ \sigma ^{2}\left( x_{1C}\right) \sigma ^{2}\left(
v_{x1C}\right) +\sigma ^{2}\left( y_{1C}\right) \sigma ^{2}\left(
v_{y1C}\right) \right] +63.5\sigma ^{2}\left( z_{1C}\right) \sigma
^{2}\left( v_{z1C}\right)  \notag \\
& +16.4\left[ \sigma ^{2}\left( x_{2C}\right) \sigma ^{2}\left(
v_{x2C}\right) +\sigma ^{2}\left( y_{2C}\right) \sigma ^{2}\left(
v_{y2C}\right) \right] +65.6\sigma ^{2}\left( z_{2C}\right) \sigma
^{2}\left( v_{z2C}\right)  \notag \\
& +478\left[ \sigma ^{4}\left( x_{1F}\right) +\sigma ^{4}\left(
y_{1F}\right) \right] +1910\sigma ^{4}\left( z_{1F}\right) +490\left[ \sigma
^{4}\left( x_{2F}\right) +\sigma ^{4}\left( y_{2F}\right) \right]
+1960\sigma ^{4}\left( z_{2F}\right) +  \notag \\
& +0.406\left[ \sigma ^{4}\left( v_{x1F}\right) +\sigma ^{4}\left(
v_{y1F}\right) \right] +1.63\sigma ^{4}\left( v_{z1F}\right) +0.439\left[
\sigma ^{4}\left( v_{x2F}\right) +\sigma ^{4}\left( v_{y2F}\right) \right]
+1.76\sigma ^{4}\left( v_{z2F}\right)  \notag \\
& +26.0\left[ \sigma ^{2}\left( x_{1F}\right) \sigma ^{2}\left(
v_{x1F}\right) +\sigma ^{2}\left( y_{1F}\right) \sigma ^{2}\left(
v_{y1F}\right) \right] +104\sigma ^{2}\left( z_{1F}\right) \sigma ^{2}\left(
v_{z1F}\right)  \notag \\
& \left. +25.8\left[ \sigma ^{2}\left( x_{2F}\right) \sigma ^{2}\left(
v_{x2F}\right) +\sigma ^{2}\left( y_{2F}\right) \sigma ^{2}\left(
v_{y2F}\right) \right] +103\sigma ^{2}\left( z_{2F}\right) \sigma ^{2}\left(
v_{z2F}\right) \right\} ^{1/2},  \tag{S.1a}  \label{S.1a} \\
s\left( \Delta _{s}^{\left( 2\right) }\phi \right) & =12.3\left[ \sigma
^{2}\left( x_{1C}\right) +\sigma ^{2}\left( y_{1C}\right) \right]
-24.7\sigma ^{2}\left( z_{1C}\right) +12.2\left[ \sigma ^{2}\left(
x_{2C}\right) +\sigma ^{2}\left( y_{2C}\right) \right] -24.3\sigma
^{2}\left( v_{z2C}\right)  \notag \\
& +0.375\left[ \sigma ^{2}\left( v_{x1C}\right) +\sigma ^{2}\left(
v_{y1C}\right) \right] -0.750\sigma ^{2}\left( v_{z1C}\right) +0.351\left[
\sigma ^{2}\left( v_{x2C}\right) +\sigma ^{2}\left( v_{y2C}\right) \right]
-0.702\sigma ^{2}\left( v_{z2C}\right)  \notag \\
& +15.5\left[ \sigma ^{2}\left( x_{1F}\right) +\sigma ^{2}\left(
y_{1F}\right) \right] -30.9\sigma ^{2}\left( z_{1F}\right) +15.7\left[
\sigma ^{2}\left( x_{2F}\right) +\sigma ^{2}\left( y_{2F}\right) \right]
-31.3\sigma ^{2}\left( z_{2F}\right)  \notag \\
& +0.451\left[ \sigma ^{2}\left( v_{x1F}\right) +\sigma ^{2}\left(
v_{y1F}\right) \right] -0.901\sigma ^{2}\left( v_{z1F}\right) +0.468\left[
\sigma ^{2}\left( v_{x2F}\right) +\sigma ^{2}\left( v_{y2F}\right) \right]
-0.937\sigma ^{2}\left( v_{z2F}\right) .  \tag{S.1b}  \label{S.1b}
\end{align}

\begin{center}
TABLE S.II: The same as Table S.I, but at the optimal choice of the atomic
positions and velocities in the C-configuration.

\begin{tabular}{|c|c|}
\hline
Term & C-configuration \\ \hline
Linear in position & $5.82\cdot 10^{-5}\delta z_{1C}-9.34\cdot 10^{-7}\delta
z_{2C}$ \\ \hline
Linear in velocity & $1.30\cdot 10^{-5}\delta v_{z1C}-1.62\cdot
10^{-7}\delta v_{z2C}$ \\ \hline
Nonlinear in position & $-17.7\left( \delta x_{1C}^{2}+\delta
y_{1C}^{2}\right) +35.35\delta z_{1C}^{2}-15.1\left( \delta
x_{2C}^{2}+\delta y_{2C}^{2}\right) +30.1\delta z_{2C}^{2}$ \\ \hline
Nonlinear in velocity & $-0.489\left( \delta v_{x1C}^{2}+\delta
v_{y1C}^{2}\right) +0.977\delta v_{z1C}^{2}-0.451\left( \delta
v_{x2C}^{2}+\delta v_{y2C}^{2}\right) +0.902\delta v_{z2C}^{2}$ \\ \hline
$%
\begin{tabular}{c}
Position-velocity \\ 
cross term%
\end{tabular}%
$ & $-5.66\left( \delta v_{x1C}\delta x_{1C}+\delta v_{y1C}\delta
y_{1C}\right) +11.3\delta v_{z1C}\delta z_{1C}-4.82\left( \delta
v_{x2C}\delta x_{2C}+\delta v_{y2C}\delta y_{2C}\right) +9.64\delta
v_{z2C}\delta z_{2C}$ \\ \hline
\end{tabular}
\end{center}

\end{subequations}
\begin{subequations}
\label{S.2}
\begin{align}
\sigma \left( \Delta _{s}^{\left( C\right) }\phi \right) & =\left\{ 625\left[
\sigma ^{4}\left( x_{1C}\right) +\sigma ^{4}\left( y_{1C}\right) \right]
+2500\sigma ^{4}\left( z_{1C}\right) +454\left[ \sigma ^{4}\left(
x_{2C}\right) +\sigma ^{4}\left( y_{2C}\right) \right] +1816\sigma
^{4}\left( z_{2C}\right) \right.  \notag \\
& +0.477\left[ \sigma ^{4}\left( v_{x1C}\right) +\sigma ^{4}\left(
v_{y1C}\right) \right] +1.91\sigma ^{4}\left( v_{z1C}\right) +0.407\left[
\sigma ^{4}\left( v_{x2C}\right) +\sigma ^{4}\left( v_{y2C}\right) \right] 
\notag \\
& +1.63\sigma ^{4}\left( v_{z2C}\right) +32.0\left[ \sigma ^{2}\left(
x_{1C}\right) \sigma ^{2}\left( v_{x1C}\right) +\sigma ^{2}\left(
y_{1C}\right) \sigma ^{2}\left( v_{y1C}\right) \right] +128\sigma ^{2}\left(
z_{1C}\right) \sigma ^{2}\left( v_{z1C}\right)  \notag \\
& \left. +23.2\left[ \sigma ^{2}\left( x_{2C}\right) \sigma ^{2}\left(
v_{x2C}\right) +\sigma ^{2}\left( y_{2C}\right) \sigma ^{2}\left(
v_{y2C}\right) \right] +93.0\sigma ^{2}\left( z_{2C}\right) \sigma
^{2}\left( v_{z2C}\right) \right\} ^{1/2},  \tag{S.2a}  \label{S.2a} \\
s\left( \Delta _{s}^{\left( C\right) }\phi \right) & =-17.7\left[ \sigma
^{2}\left( x_{1C}\right) +\sigma ^{2}\left( y_{1C}\right) \right]
+35.3\sigma ^{2}\left( z_{1C}\right) -15.1\left[ \sigma ^{2}\left(
x_{2C}\right) +\sigma ^{2}\left( y_{2C}\right) \right] +30.1\sigma
^{2}\left( v_{z2C}\right)  \notag \\
& -0.489\left[ \sigma ^{2}\left( v_{x1C}\right) +\sigma ^{2}\left(
v_{y1C}\right) \right] +0.977\sigma ^{2}\left( v_{z1C}\right) -0.451\left[
\sigma ^{2}\left( v_{x2C}\right) +\sigma ^{2}\left( v_{y2C}\right) \right]
+0.902\sigma ^{2}\left( v_{z2C}\right) .  \tag{S.2b}  \label{S.2b}
\end{align}

\begin{center}
TABLE S.III: The same as in Table S.I, but at the optimal choice of the
atomic positions and velocities in the $C-$configuration, and after
horizontal moving apart the halves of the source mass in the opposite
directions by the distance (\ref{29}) in the $F-$configuration.

\begin{tabular}{|c|c|c|}
\hline
Term & C-configuration & F-configuration \\ \hline
$\Delta _{s}^{\left( I\right) }\phi /\Delta ^{\left( 2\right) }\phi $ & $%
1.020272$ & $-0.020272$ \\ \hline
Linear in position & $5.94\cdot 10^{-5}\delta z_{1C}-9.53\cdot 10^{-7}\delta
z_{2C}$ & $\left[ 4.51\delta z_{1F}-4.55\delta z_{2F}\right] \cdot 10^{-2}$
\\ \hline
Linear in velocity & $1.33\cdot 10^{-5}\delta v_{z1C}-1.66\cdot
10^{-7}\delta v_{z2C}$ & $\left[ 7.21\delta v_{z1F}-7.28\delta v_{z2F}\right]
\cdot 10^{-3}$ \\ \hline
Nonlinear in position & $%
\begin{array}{c}
-18.0\left( \delta x_{1C}^{2}+\delta y_{1C}^{2}\right) +36.1\delta z_{1C}^{2}
\\ 
-15.4\left( \delta x_{2C}^{2}+\delta y_{2C}^{2}\right) +30.7\delta z_{2C}^{2}%
\end{array}%
$ & $%
\begin{array}{c}
\left[ -72.6\delta x_{1F}^{2}+19.2\delta y_{1F}^{2}+53.4\delta
z_{1F}^{2}\right. \\ 
-9.93\delta x_{1F}\delta v_{y1F} \\ 
-71.5\delta x_{2F}^{2}+18.8\delta y_{2F}^{2}-52.6\delta z_{2F}^{2} \\ 
\left. -9.78\delta x_{2F}\delta y_{2F}\right] \cdot 10^{-3}%
\end{array}%
$ \\ \hline
Nonlinear in velocity & $%
\begin{array}{c}
-0.498\left( \delta v_{x1C}^{2}+\delta v_{y1C}^{2}\right) \\ 
+0.997\delta v_{z1C}^{2} \\ 
-0.460\left( \delta v_{x2C}^{2}+\delta v_{y2C}^{2}\right) \\ 
+0.921\delta v_{z2C}^{2}%
\end{array}%
$ & $%
\begin{array}{c}
\left[ -21.9\delta v_{x1F}^{2}+5.81\delta v_{y1F}^{2}\right. \\ 
+16.1\delta v_{z1F}^{2}-3.00\delta v_{x1F}\delta v_{y1F} \\ 
-21.0\delta v_{x2F}^{2}+5.52\delta v_{y2F}^{2} \\ 
\left. +15.5\delta v_{z2F}^{2}-2.87\delta v_{x2F}\delta v_{y2F}\right] \cdot
10^{-4}%
\end{array}%
$ \\ \hline
position-velocity cross term & $%
\begin{array}{c}
-5.77\left( \delta v_{x1C}\delta x_{1C}+\delta v_{y1C}\delta y_{1C}\right)
\\ 
+11.5\delta v_{z1C}\delta z_{1C} \\ 
-4.92\left( \delta v_{x2C}\delta x_{2C}+\delta v_{y2C}\delta y_{2C}\right)
\\ 
+9.84\delta v_{z2C}\delta z_{2C}%
\end{array}%
$ & $%
\begin{array}{c}
\left[ -23.2\delta v_{x1F}\delta x_{1F}+6.13\delta v_{y1F}\delta
y_{1F}\right. \\ 
+17.1\delta v_{z1F}\delta z_{1F}-1.59\left( \delta v_{y1F}\delta
x_{1F}+\delta v_{x1F}\delta y_{1F}\right) \\ 
-22.9\delta v_{x2F}\delta x_{2F}+6.02\delta v_{y2F}\delta y_{2F}+16.8\delta
v_{z2F}\delta z_{2F} \\ 
\left. -1.56\left( \delta v_{y2F}\delta x_{2F}+\delta v_{x2F}\delta
y_{2F}\right) \right] \cdot 10^{-3}%
\end{array}%
$ \\ \hline
\end{tabular}
\end{center}

\end{subequations}
\begin{subequations}
\label{S.3}
\begin{align}
\sigma \left( \Delta _{s}^{\left( 2\right) }\phi \right) & =\left\{ 650. 
\left[ \sigma ^{4}\left( x_{1C}\right) +\sigma ^{4}\left( y_{1C}\right) %
\right] +2600\sigma ^{4}\left( z_{1C}\right) +473\left[ \sigma ^{4}\left(
x_{2C}\right) +\sigma ^{4}\left( y_{2C}\right) \right] +1890\sigma
^{4}\left( z_{2C}\right) \right.  \notag \\
& +0.497\left[ \sigma ^{4}\left( v_{x1C}\right) +\sigma ^{4}\left(
v_{y1C}\right) \right] +1.99\sigma ^{4}\left( v_{z1C}\right) +0.424\left[
\sigma ^{4}\left( v_{x2C}\right) +\sigma ^{4}\left( v_{y2C}\right) \right]
+1.69\sigma ^{4}\left( v_{z2C}\right)  \notag \\
& +33.3\left[ \sigma ^{2}\left( x_{1C}\right) \sigma ^{2}\left(
v_{x1C}\right) +\sigma ^{2}\left( y_{1C}\right) \sigma ^{2}\left(
v_{y1C}\right) \right] +133\sigma ^{2}\left( z_{1C}\right) \sigma ^{2}\left(
v_{z1C}\right)  \notag \\
& +24.2\left[ \sigma ^{2}\left( x_{2C}\right) \sigma ^{2}\left(
v_{x2C}\right) +\sigma ^{2}\left( y_{2C}\right) \sigma ^{2}\left(
v_{y2C}\right) \right] +96.8\sigma ^{2}\left( z_{2C}\right) \sigma
^{2}\left( v_{z2C}\right)  \notag \\
& +\left[ 2.03\sigma ^{2}\left( z_{1F}\right) +2.07\sigma ^{2}\left(
z_{2F}\right) \right] \cdot 10^{-3}+\left[ 5.20\sigma ^{2}\left(
v_{z1F}\right) +5.30\sigma ^{2}\left( v_{z2F}\right) \right] \cdot 10^{-5} 
\notag \\
& +\left[ 1050\sigma ^{4}\left( x_{1F}\right) +73.4\sigma ^{4}\left(
y_{1F}\right) +9.85\sigma ^{2}\left( x_{1F}\right) \sigma ^{2}\left(
y_{1F}\right) +570\sigma ^{4}\left( z_{1F}\right) \right.  \notag \\
& \left. +1020\sigma ^{4}\left( x_{2F}\right) +70.9\sigma ^{4}\left(
y_{2F}\right) +9.56\sigma ^{2}\left( x_{2F}\right) \sigma ^{2}\left(
y_{2F}\right) +554\sigma ^{4}\left( z_{2F}\right) \right] \cdot 10^{-5} 
\notag \\
& +\left[ 963\sigma ^{4}\left( v_{x1F}\right) +67.5\sigma ^{4}\left(
v_{y1F}\right) +9.00\sigma ^{2}\left( v_{x1F}\right) \sigma ^{2}\left(
v_{y1F}\right) +520\sigma ^{4}\left( v_{z1F}\right) \right.  \notag \\
& \left. +881\sigma ^{4}\left( v_{x2F}\right) +60.9\sigma ^{4}\left(
v_{y2F}\right) +8.25\sigma ^{2}\left( v_{x2F}\right) \sigma ^{2}\left(
v_{y2F}\right) +479\sigma ^{4}\left( v_{z2F}\right) \right] \cdot 10^{-8} 
\notag \\
& +\left[ 539\sigma ^{2}\left( x_{1F}\right) \sigma ^{2}\left(
v_{x1F}\right) +2.52\sigma ^{2}\left( x_{1F}\right) \sigma ^{2}\left(
v_{y1F}\right) +37.6\sigma ^{2}\left( y_{1F}\right) \sigma ^{2}\left(
v_{y1F}\right) \right.  \notag \\
& +292\sigma ^{2}\left( z_{1F}\right) \sigma ^{2}\left( v_{z1F}\right)
+523\sigma ^{2}\left( x_{2F}\right) \sigma ^{2}\left( v_{x2F}\right)
+36.3\sigma ^{2}\left( y_{2F}\right) \sigma ^{2}\left( v_{y2F}\right)  \notag
\\
& \left. \left. +2.45\sigma ^{2}\left( y_{2F}\right) \sigma ^{2}\left(
v_{x2F}\right) +284\sigma ^{2}\left( z_{2F}\right) \sigma ^{2}\left(
v_{z2F}\right) \right] \cdot 10^{-6}\right\} ^{1/2},  \tag{S.3a}
\label{S.3a} \\
s\left( \Delta _{s}^{\left( 2\right) }\phi \right) & =-18.0\left[ \sigma
^{2}\left( x_{1C}\right) +\sigma ^{2}\left( y_{1C}\right) \right]
+36.1\sigma ^{2}\left( z_{1C}\right) -15.4\left[ \sigma ^{2}\left(
x_{2C}\right) +\sigma ^{2}\left( y_{2C}\right) \right] +30.74\sigma
^{2}\left( v_{z2C}\right)  \notag \\
& -0.498\left[ \sigma ^{2}\left( v_{x1C}\right) +\sigma ^{2}\left(
v_{y1C}\right) \right] +0.997\sigma ^{2}\left( v_{z1C}\right) -0.460\left[
\sigma ^{2}\left( v_{x2C}\right) +\sigma ^{2}\left( v_{y2C}\right) \right]
+0.921\sigma ^{2}\left( v_{z2C}\right)  \notag \\
& \left[ -726\sigma ^{2}\left( x_{1F}\right) +192\sigma ^{2}\left(
y_{1F}\right) +534\sigma ^{2}\left( z_{1F}\right) -715\sigma ^{2}\left(
x_{2F}\right) +188\sigma ^{2}\left( y_{2F}\right) +526\sigma ^{2}\left(
z_{2F}\right) -21.9\sigma ^{2}\left( v_{x1F}\right) \right.  \notag \\
& \left. +5.81\sigma ^{2}\left( v_{y1F}\right) +16.1\sigma ^{2}\left(
v_{z1F}\right) -21.0\sigma ^{2}\left( v_{x2F}\right) +5.52\sigma ^{2}\left(
v_{y2F}\right) +15.5\sigma ^{2}\left( v_{z2F}\right) \right] \cdot 10^{-4}. 
\tag{S.3b}  \label{S.3b}
\end{align}

\begin{center}
TABLE S.IV: The same as Table S.II, but for the 630 culinders.

\begin{tabular}{|c|c|}
\hline
Term & C-configuration \\ \hline
Linear in position & $-9.25\cdot 10^{-8}\delta z_{1C}+1.28\cdot
10^{-7}\delta z_{2C}$ \\ \hline
Linear in velocity & $-1.79\cdot 10^{-8}\delta v_{z1C}+1.18\ast
10^{-7}\delta v_{z2C}$ \\ \hline
Nonlinear in position & $4.92\left( \delta x_{1C}^{2}+\delta
y_{1C}^{2}\right) -9.84\delta z_{1C}^{2}+2.97\left( \delta x_{2C}^{2}+\delta
y_{2C}^{2}\right) -5.95\delta z_{2C}^{2}$ \\ \hline
Nonlinear in velocity & $0.306\left( \delta v_{x1C}^{2}+\delta
v_{y1C}^{2}\right) -0.612\delta v_{z1C}^{2}+0.197\left( \delta
v_{x2C}^{2}+\delta v_{y2C}^{2}\right) -0.393\delta v_{z2C}^{2}$ \\ \hline
$%
\begin{tabular}{c}
Position-velocity \\ 
cross term%
\end{tabular}%
$ & $2.39\left( \delta v_{x1C}\delta x_{1C}+\delta v_{y1C}\delta
y_{1C}\right) -4.78\delta v_{z1C}\delta z_{1C}+1.45\left( \delta
v_{x2C}\delta x_{2C}+\delta v_{y2C}\delta y_{2C}\right) -2.89\delta
v_{z2C}\delta z_{2C}$ \\ \hline
\end{tabular}
\end{center}

\end{subequations}
\begin{subequations}
\label{S.4}
\begin{align}
\sigma \left( \Delta _{s}^{\left( C\right) }\phi \right) & =\left\{ 48.4 
\left[ \sigma ^{4}\left( x_{1C}\right) +\sigma ^{4}\left( y_{1C}\right) %
\right] +193\sigma ^{4}\left( z_{1C}\right) +17.7\left[ \sigma ^{4}\left(
x_{2C}\right) +\sigma ^{4}\left( y_{2C}\right) \right] +70.8\sigma
^{4}\left( z_{2C}\right) \right.  \notag \\
& +0.187\left[ \sigma ^{4}\left( v_{x1C}\right) +\sigma ^{4}\left(
v_{y1C}\right) \right] +0.749\sigma ^{4}\left( v_{z1C}\right) +0.0773\left[
\sigma ^{4}\left( v_{x2C}\right) +\sigma ^{4}\left( v_{y2C}\right) \right]
+0.309\sigma ^{4}\left( v_{z2C}\right)  \notag \\
& +5.71\left[ \sigma ^{2}\left( x_{1C}\right) \sigma ^{2}\left(
v_{x1C}\right) +\sigma ^{2}\left( y_{1C}\right) \sigma ^{2}\left(
v_{y1C}\right) \right] +22.8\sigma ^{2}\left( z_{1C}\right) \sigma
^{2}\left( v_{z1C}\right)  \notag \\
& +\left. 2.09\left[ \sigma ^{2}\left( x_{2C}\right) \sigma ^{2}\left(
v_{x2C}\right) +\sigma ^{2}\left( y_{2C}\right) \sigma ^{2}\left(
v_{y2C}\right) \right] +8.36\sigma ^{2}\left( z_{2C}\right) \sigma
^{2}\left( v_{z2C}\right) \right\} ^{1/2},  \tag{S.4a}  \label{S.4a} \\
s\left( \Delta _{s}^{\left( C\right) }\phi \right) & =4.92\left[ \sigma
^{2}\left( x_{1C}\right) +\sigma ^{2}\left( y_{1C}\right) \right]
-9.84\sigma ^{2}\left( z_{1C}\right) +2.97\left[ \sigma ^{2}\left(
x_{2C}\right) +\sigma ^{2}\left( y_{2C}\right) \right] -5.95\sigma
^{2}\left( v_{z2C}\right)  \notag \\
& +0.306\left[ \sigma ^{2}\left( v_{x1C}\right) +\sigma ^{2}\left(
v_{y1C}\right) \right] -0.612\sigma ^{2}\left( v_{z1C}\right) +0.197\left[
\sigma ^{2}\left( v_{x2C}\right) +\sigma ^{2}\left( v_{y2C}\right) \right]
-0.393\sigma ^{2}\left( v_{z2C}\right) ,  \tag{S.4b}  \label{S.4b}
\end{align}

\begin{center}
TABLE S.V: The same as Table S.III, but for the 630 cylinders.

\begin{tabular}{|c|c|c|}
\hline
Term & C-configuration & F-configuration \\ \hline
$\phi _{s}^{\left( I\right) }/\Delta ^{\left( 2\right) }\phi $ & $1.042989$
& $-0.042989$ \\ \hline
Linear in position & $-9.65\cdot 10^{-8}\delta z_{1C}+1.33\cdot
10^{-7}\delta z_{2C}$ & $\left( 3.30\delta z_{1F}-3.41\delta z_{2F}\right)
\cdot 10^{-2}$ \\ \hline
Linear in velocity & $-1.86\cdot 10^{-8}\delta v_{z1C}+1.23\cdot
10^{-7}\delta v_{z2C}$ & $\left( 8.03\delta v_{z1F}-8.28\delta
v_{z2F}\right) \cdot 10^{-3}$ \\ \hline
Nonlinear in position & $%
\begin{array}{c}
5.13\left( \delta x_{1C}^{2}+\delta y_{1C}^{2}\right) -10.3\delta z_{1C}^{2}
\\ 
+3.10\left( \delta x_{2C}^{2}+\delta y_{2C}^{2}\right) -6.21\delta z_{2C}^{2}%
\end{array}%
$ & $%
\begin{array}{c}
\left( -33.5\delta x_{1F}^{2}+9.01\delta y_{1F}^{2}-1.63\delta x_{1F}\delta
y_{1F}\right. \\ 
+24.5\delta z_{1F}^{2}-32.7\delta x_{2F}^{2}+8.75\delta y_{2F}^{2} \\ 
\left. -1.60\delta x_{2F}\delta y_{2F}+24.0\delta z_{2F}^{2}\right) \cdot
10^{-3}%
\end{array}%
$ \\ \hline
Nonlinear in velocity & $%
\begin{array}{c}
0.319\left( \delta v_{x1C}^{2}+\delta v_{y1C}^{2}\right) -0.638\delta
v_{z1C}^{2} \\ 
0.205\left( \delta v_{x2C}^{2}+\delta v_{y2C}^{2}\right) -0.410\delta
v_{z2C}^{2}%
\end{array}%
$ & $%
\begin{array}{c}
\left( 23.2\delta v_{x1F}^{2}+6.28\delta v_{y1F}^{2}+17.0\delta
v_{z1F}^{2}\right. \\ 
+1.13\delta v_{x1F}\delta v_{y1F}+22.3\delta v_{x2F}^{2}+5.94\delta
v_{y2F}^{2} \\ 
\left. +1.09\delta v_{x2F}\delta v_{y2F}+16.3\delta v_{z2F}^{2}\right) \cdot
10^{-4}%
\end{array}%
$ \\ \hline
Position-velocity cross term & $%
\begin{array}{c}
2.49\left( \delta v_{x1C}\delta x_{1C}+\delta v_{y1C}\delta y_{1C}\right) \\ 
-4.99\delta v_{z1C}\delta z_{1C} \\ 
+1.51\left( \delta v_{x2C}\delta x_{2C}+\delta v_{y2C}\delta y_{2C}\right)
\\ 
-3.02\delta v_{z2C}\delta z_{2C}%
\end{array}%
$ & $%
\begin{array}{c}
\left[ -163\delta v_{x1F}\delta x_{1F}+43.8\delta v_{y1F}\delta y_{1F}\right.
\\ 
-3.97\left( \delta x_{1F}\delta v_{y1F}+\delta y_{1F}\delta v_{x1F}\right)
\\ 
+119\delta v_{z1F}\delta z_{1F}-159\delta v_{x2F}\delta x_{2F} \\ 
+42.5\delta v_{y2F}\delta y_{2F} \\ 
-3.88\left( \delta v_{x2F}\delta y_{2F}+\delta v_{y2F}\delta x_{2F}\right)
\\ 
\left. +117\delta v_{z2F}\delta z_{2F}\right] \cdot 10^{-4}%
\end{array}%
$ \\ \hline
\end{tabular}
\end{center}

\end{subequations}
\begin{subequations}
\label{S.5}
\begin{align}
\sigma \left( \Delta _{s}^{\left( 2\right) }\phi \right) & =\left\{ 52.6 
\left[ \sigma ^{4}\left( x_{1C}\right) +\sigma ^{4}\left( y_{1C}\right) %
\right] +210\sigma ^{4}\left( z_{1C}\right) +19.3\left[ \sigma ^{4}\left(
x_{2C}\right) +\sigma ^{4}\left( y_{2C}\right) \right] +77.0\sigma
^{4}\left( z_{2C}\right) \right.  \notag \\
& +0.204\left[ \sigma ^{4}\left( v_{x1C}\right) +\sigma ^{4}\left(
v_{y1C}\right) \right] +0.814\sigma ^{4}\left( v_{z1C}\right) +0.0841\left[
\sigma ^{4}\left( v_{x2C}\right) +\sigma ^{4}\left( v_{y2C}\right) \right] 
\notag \\
& +0.336\sigma ^{4}\left( v_{z2C}\right) +6.21\left[ \sigma ^{2}\left(
x_{1C}\right) \sigma ^{2}\left( v_{x1C}\right) +\sigma ^{2}\left(
y_{1C}\right) \sigma ^{2}\left( v_{y1C}\right) \right] +24.9\sigma
^{2}\left( z_{1C}\right) \sigma ^{2}\left( v_{z1C}\right)  \notag \\
& +2.27\left[ \sigma ^{2}\left( x_{2C}\right) \sigma ^{2}\left(
v_{x2C}\right) +\sigma ^{2}\left( y_{2C}\right) \sigma ^{2}\left(
v_{y2C}\right) \right] +9.10\sigma ^{2}\left( z_{2C}\right) \sigma
^{2}\left( v_{z2C}\right)  \notag \\
& +\left[ 1.09\sigma ^{2}\left( z_{1F}\right) +1.16\sigma ^{2}\left(
z_{2F}\right) \right] \cdot 10^{-3}+\left[ 6.44\sigma ^{2}\left(
v_{z1F}\right) +6.85\sigma ^{2}\left( v_{z2F}\right) \right] \cdot 10^{-5} 
\notag \\
& +\left[ 2240\sigma ^{4}\left( x_{1F}\right) +162\sigma ^{4}\left(
y_{1F}\right) +2.66\sigma ^{2}\left( x_{1F}\right) \sigma ^{2}\left(
y_{1F}\right) +1200\sigma ^{4}\left( z_{1F}\right) \right.  \notag \\
& \left. +2140\sigma ^{4}\left( x_{2F}\right) +153\sigma ^{4}\left(
y_{2F}\right) +2.55\sigma ^{2}\left( x_{2F}\right) \sigma ^{2}\left(
y_{2F}\right) +1150\sigma ^{4}\left( z_{2F}\right) \right] \cdot 10^{-6} 
\notag \\
& +\left[ 1080\sigma ^{4}\left( v_{x1F}\right) +78.8\sigma ^{4}\left(
v_{y1F}\right) +1.28\sigma ^{2}\left( v_{x1F}\right) \sigma ^{2}\left(
v_{y1F}\right) +575\sigma ^{4}\left( v_{z1F}\right) \right.  \notag \\
& \left. +993\sigma ^{4}\left( v_{x2F}\right) +70.6\sigma ^{4}\left(
v_{y2F}\right) +1.18\sigma ^{2}\left( v_{x2F}\right) \sigma ^{2}\left(
v_{y2F}\right) +534\sigma ^{4}\left( v_{z2F}\right) \right] \cdot 10^{-8} 
\notag \\
& +\left[ 2650\sigma ^{2}\left( x_{1F}\right) \sigma ^{2}\left(
v_{x1F}\right) +192\sigma ^{2}\left( y_{1F}\right) \sigma ^{2}\left(
v_{y1F}\right) +1.57\left( \sigma ^{2}\left( y_{1F}\right) \sigma ^{2}\left(
v_{x1F}\right) +\sigma ^{2}\left( x_{1F}\right) \sigma ^{2}\left(
v_{y1F}\right) \right) \right.  \notag \\
& +1410\sigma ^{2}\left( z_{1F}\right) \sigma ^{2}\left( v_{z1F}\right)
+2530\sigma ^{2}\left( x_{2F}\right) \sigma ^{2}\left( v_{x2F}\right)
+181\sigma ^{2}\left( y_{2F}\right) \sigma ^{2}\left( v_{y2F}\right)  \notag
\\
& +\left. \left. 1.51\left( \sigma ^{2}\left( x_{2F}\right) \sigma
^{2}\left( v_{y2F}\right) +\sigma ^{2}\left( y_{2F}\right) \sigma ^{2}\left(
v_{x2F}\right) \right) +1360\sigma ^{2}\left( z_{2F}\right) \sigma
^{2}\left( v_{z2F}\right) \right] \cdot 10^{-7}\right\} ^{1/2},  \tag{S.5a}
\label{S.5a} \\
s\left( \Delta _{s}^{\left( 2\right) }\phi \right) & =5.13\left[ \sigma
^{2}\left( x_{1C}\right) +\sigma ^{2}\left( y_{1C}\right) \right]
-10.2\sigma ^{2}\left( z_{1C}\right) +3.10\left[ \sigma ^{2}\left(
x_{2C}\right) +\sigma ^{2}\left( y_{2C}\right) \right] -6.22\sigma
^{2}\left( z_{2C}\right)  \notag \\
& +0.319\left[ \sigma ^{2}\left( v_{x1C}\right) +\sigma ^{2}\left(
v_{y1C}\right) \right] -0.638\sigma ^{2}\left( v_{z1C}\right) +0.205\left[
\sigma ^{2}\left( v_{x2C}\right) +\sigma ^{2}\left( v_{y2C}\right) \right]
-0.410\sigma ^{2}\left( v_{z2C}\right)  \notag \\
& +\left[ -33.5\sigma ^{2}\left( x_{1F}\right) +9.01\sigma ^{2}\left(
y_{1F}\right) +24.5\sigma ^{2}\left( z_{1F}\right) -32.7\sigma ^{2}\left(
x_{2F}\right) +8.75\sigma ^{2}\left( y_{2F}\right) +24.0\sigma ^{2}\left(
z_{2F}\right) \right] \cdot 10^{-3}  \notag \\
& +\left[ -23.2\sigma ^{2}\left( v_{x1F}\right) +6.28\sigma ^{2}\left(
v_{y1F}\right) +17.0\sigma ^{2}\left( v_{z1F}\right) -22.3\sigma ^{2}\left(
v_{x2F}\right) +5.94\sigma ^{2}\left( v_{y2F}\right) +16.3\sigma ^{2}\left(
v_{z2F}\right) \right] \cdot 10^{-4}  \tag{S.5b}  \label{S.5b}
\end{align}

\section{Manuscript consideration}

\subsection{Date: Tuesday, July 16, 2019 8:53:52 AM}

From: pra@aps.org

To: bdubetsky@gmail.com

Subject: Your\_manuscript AG11947 Dubetsky

Re: AG11947

Optimization of the atom interferometer phase produced by the set of
cylindrical source masses to measure the Newtonian gravity constant by B.
Dubetsky

Dear Dr. Dubetsky,

The Physical Review editors attempt to accept only papers that are
scientifically sound, important to the field, and contain significant new
results in physics. We judge that these acceptance criteria are not met by
your manuscript.

We regret that consequently we cannot accept the paper for publication in
the Physical Review.

Yours sincerely,

Doerte Blume

Associate Editor

\subsection{Date: Friday, September 6, 2019 7:12:08 PM}

From: Boris Dubetsky

To: pra@aps.org

Subject: Re: Your\_manuscript AG11947 Dubetsky

Attachments: image.png

cylinders3\_short..pdf

Dear Dr. Blume,

I examined the Physical Review A criteria before I dared to submit my
article. They are strict, but they are not quantitative. You wrote about two
parameters: scientific importance to the field and significant new results
in physics. My manuscript AG11947 is devoted to the important scientific
problem, measurement of the Newton gravity constant G. which is measured
worse than all other fundamental constant. It also contains new results:
restriction of the measurement accuracy, caused by the gradient of the Earth
gravity field, the necessity to change the approach to the error budget,
caused by the point that measurements have been ALWAISE performed near
extremal values of the atomic position and velocity, the new systematic
error arising from the same reason, inventing of the regime, where errors
caused by the atomic finite temperature and finite size of atomic clouds
disappear.

I see only one way to estimate scientific importance and significance of the
new results fairly - by comparing my article with others published in your
Journal. Please, understand me correctly, I am not complaining, not
claiming, why did you accept those articles and reject my one. I indeed
don't know another way to estimate of my article. If you are aware of
another way, please let me know it.

Last two theoretical articles devoted to the atom interferometry are A.
Bertoldi et.al, Phys. Rev. A 99, 033619 (2019), Ya-Lie Wang et.al, Phys.
Rev. A 98, 053604 (2018). Both articles consider only one systematic error
caused by the Raman pulses finite duration. I consider the whole class of
the errors caused by the use of extreme variables and show that one has to
change totally the approach of including those variables in the error
budget. I considered both systematic and statistical errors. If statistical
errors were considered previously in [2-4, 14, 18] just incorrectly, then
systematic errors were totally missed.

It was accepted in the articles [2,3] that the Earth gravity field plays no
role in the G measurement. I showed that it is incorrect. Earth gravity
field gradient does affect on the error budget. Now look, in the articles
[2,3] published in the journals Science and Nature it was missed part of the
error budget, which is larger than all other parts, I found it, but
according to your conclusion, for the journal Physical Review A it is
insignificant.

Would you please to reconsider your conclusion about the manuscript.

I prepared a new version of the article. To emphasize the significance of my
findings I've changed the abstract and add the following paragraph on page 2:

"We think that the restriction (8) also played a certain role in the
experiments [2,18] and was one of the reasons for the low accuracy of the
measurement of $G$. In the article [2] for the following values of
parameters $\Delta _{s}^{2}\phi \approx 0.1$rad$,$ $k=1.47\cdot 10^{7}$m$^{%
\text{-1}},$ $\left\vert \Gamma _{Ezz}\right\vert =2.93\cdot 10^{-6}s^{\text{%
-2}}$ [16], $\sigma \left( z_{jI}\right) \thicksim 3\cdot 10^{-4}$m, $\sigma
\left( v_{z_{jI}}\right) \thicksim 2\cdot 10^{-3}$m/s, $T=150$ms one gets
from (8) $\sigma _{r}\left( \Delta ^{2}\phi \right) >6\cdot 10^{-3}$. On\
the order of magnitude, this restriction is close to the measurement
accuracy $4\cdot 10^{-3}$ in [2]. In the work [18] at $\Delta _{s}^{2}\phi
=.503$rad, $\sigma \left( z_{1I}\right) \thicksim 1.2\cdot 10^{-4}$m, $%
\sigma \left( z_{2I}\right) \thicksim 1.3\cdot 10^{-4}$m, $\sigma \left(
v_{z_{jI}}\right) \thicksim 5\cdot 10^{-3}$m/s, $\left\vert \Gamma
_{Ezz}\right\vert =3.11\cdot 10^{-6}$s$^{\text{-2}},$ $k\approx 1.61\cdot
10^{7}$m$^{\text{-1}}$, $T=150$ms one finds $\sigma _{r}\left( \Delta
^{2}\phi \right) >2.41\cdot 10^{-3}.$ This restriction 2.8 times larger than
the RSD in the error $\phi _{s}^{\left( C\right) }$ in Table 5.1, but it an
order of magnitude smaller than the accuracy of the measurement of $G$ in
[18]. It should be emphasized that the influence of the gradient of Earth
gravity field on the accuracy of the measurement of the Newton gravitational
constant was not discussed clearly in all preceding articles [2-4,14,18]."

To make the manuscript better readable, I withdraw tables and lengthy
expressions with numerical data, but I saved references on those tables and
equations, which are still presented in the online version of the

article arXiv:1907.03352v3 [physics.atom-ph].

Please, find enclosed the new version.

Yours sincerely,

Boris Dubetsky

\subsection{Date: Wednesday, September 18, 2019 11:04:07 PM}

From: Boris Dubetsky

To: pra@aps.org

Subject: Re: Publication Rights AG11947 Dubetsky

Attachments: cylinders4\_short.pdf

Dear Physical Review A Journal Services representative,

I am really sorry, but unfortunately, I found an error in my calculations.
The error happened owing to the wrong assumption that contributions to the
atom interferometer phase produced by the Earth field and source mass field
are statistically independent. After calculations have been corrected, I
have changed Abstract, Sections I, IV, and VII. Would you please, find
enclosed the manuscript after those changes. If it is not impossible, please
resend manuscript to Referee.

Sincerely yours,

Boris Dubetsky

\subsection{Date: Thursday, October 24, 2019 11:07:54 AM}

From: pra@aps.org

To: bdubetsky@gmail.com

Subject: Your\_manuscript AG11947 Dubetsky

Re: AG11947

Optimization of the atom interferometer phase produced by the set of

cylindrical source masses to measure the Newtonian gravity constant

by B. Dubetsky

Dear Dr. Dubetsky,

The above manuscript has been reviewed by two of our referees.

Comments from the reports appear below.

These comments suggest that the present manuscript is not suitable for
publication in the Physical Review.

Yours sincerely,

Xiangyu Yin

Assistant Editor

----------------------------------------------------------------------

Report of the First Referee -- AG11947/Dubetsky

----------------------------------------------------------------------

In this manuscript the author presents a couple of results:

First, directly quoting the abstract, "An analytical expression for the
gravitational field of a homogeneous cylinder is derived."

Second, the results of two experiments, one completed in 2014, the other,
presently at the proposal stage, both measuring the gravitational constant G
with cold atom interferometers and cylindrical source masses, are carefully
analyzed in order to optimize and quantify the expected uncertainty on G as
a function of the source masses' positions.

I think that the material is potentially interesting but in its present form
it cannot be considered for publication. The note [28] states:

"Tables and lengthy equations containing coefficients obtained numerically
one can find only in the online version of this article B. Dubetsky
arXiv:1907.03352v4 [physics.atom-ph]. Starting from the Sec. IV the
references on those tables and equations are put in braces."

The article submitted is then not complete so, i.e., tables \{I\} to
\{V\}cited in the text are nowhere to be found in the manuscript and are
available only on arXiv. Phys. Rev. A offers the possibility to include
supplementary contents. The author should take advantage of this opportunity
and submit a full, self-contained, manuscript.

Since this is a matter that, I assume, can be easily managed by the author,
I went through the reading anyway, hoping to save some time, eventually, to
the review process. I have two main comments and some minor ones.

1) I checked and I have found a previous article, recently published
actually, reporting the gravitational field of a cylinder:

"Closed-form expressions for the non-axial component of the gravitational
field of an arbitrary cylinder segment" by David Miles Journal of Applied
Geophysics 159 (2018) 621--630 see
https://doi.org/10.1016/j.jappgeo.2018.10.006

Apparently, according to a reference in it, also the axial component has
been published before. See Nabighian, N., 1962. "The Gravitational
attraction of a vertical circular cylinder at points external to it."

Geofis. Pura e Appl. 53, 45--51.

2) The author correctly points out that when a function F of a random
variable X for which mu=E[X] (here E[] denotes he expectation value), has
maximum or a minimum in F(mu) then E[F(X)] is not, approximately, F(mu) but
there is a substantial bias.

Then he proceeds to evaluate the correction, assuming a known \ probability
density p(X) for X, and eventually suggests, in the conclusions, close to
eq. (50), that a shift of the order of -200 ppm should be applied to the
value of G published in [3].

A less elegant but just as accurate procedure to obtain the unbiased value
for E[F(X)] is to run a Montecarlo i.e. by sampling X over p(X) and
evaluating an average of F(X). This is actually the procedure used in [3],
as stated in the Methods section, so I think that the correction mentioned
proposed by the author is not needed.

Minor points

- Some choices of numerical parameters seem peculiar. Can for example the
author motivate the choice of "a distance between the floors of 0.20011 m"
(why the extra 110 um?) or "temperature 115 pK and radius of the atom cloud
170 um (which are larger than those observed in [7])" (why then not use then
the numbers given in [7] in an estimate).

- Choosing a given notation is a subjective matter so I take it as perfectly
fine if the author does not agree with my comments, but I would recommend to
simplify some formulas in order to improve readability. As an example
consider eq.(20) and eq. (21a-e):

1) hbar k/M is the recoil velocity so v\_r/2 = hbar k/2M could be defined
and used. By the way, to erase the Earth's gradient k should change at every
pulse. Which k should be taken here? Please clarify.

2) \TEXTsymbol{\backslash}delta g\_z, \TEXTsymbol{\backslash}gamma\_\{zm\}, 
\TEXTsymbol{\backslash}chi\_\{zmn\} are meant to be evaluated in the
integrands along two trajectories in space

a(t)= x+ v*(T\_1 + T + t) + g/2*(T\_1 + T + t)\symbol{94}2 +v\_r/2*(T + t)

b(t)= x+ v*(T\_1 + t) + g/2*(T\_1 + t)\symbol{94}2 +v\_r/2*t

Does that really need to be repeated explicitly every time? Also I find
misleading to use f[x(t)] instead of f(x(t)) i.e. using [] to indicate the
argument of a function. It is easy to assume that it is a multiplicative
factor.

Other scattered examples: why the double difference in eq.\symbol{126}(2) is 
\TEXTsymbol{\backslash}Delta\symbol{94}2 and not, i.e. \TEXTsymbol{\backslash%
}Delta\_2, so it won't be confused with a square? Why the vertical component
of Earth gradient is \TEXTsymbol{\backslash}Gamma\_\{E33\} while for the
source masses \TEXTsymbol{\backslash}gamma\_\{zm\} is use instead of,
following the first notation, with something like \TEXTsymbol{\backslash}%
Gamma\_\{S3m\}?

- eq(7) assumes that the covariance matrix is diagonal i.e. that
\{q\_1..q\_n\} are statistically independent. It should be explicitly stated.

- the moments in eq. (15a,b) are only about positions. Mentioning that the
obvious extensions for velocities and mixed terms could be useful. Maybe
just indicate with q\_i either a position x\_i or a velocity v\_i.

- Eq. (17a) is a general result however samples of cold atoms from a MOT are
usually modeled with Gaussians distributions for velocities while for
positions the density distributions are either Gaussians (at long times
after release from the MOT) or uniform ellipsoids at short times. I will not
discuss here what "long" and "short" mean in this contest. I just want to
point out that taking \TEXTsymbol{\backslash}kappa=0 should be a good
approximation for most experimental cases.

- Fig. 1: Using yellow for the trajectories is not the best choice for
contrast.

- at page 8 the author writes "..up to the third digit, they coincide with
the velocity of the atomic fountain [29] v = gT". If I got it right, this
means that v\_z\symbol{126}0 will hold at the second pulse. This situation
is always avoided in Raman interferometers because it cannot select the
direction in which momentum is absorbed i.e. +v\_r and -v\_r recoils are
equally probable, so half of the signal is lost. Recently, however, a
proposal for implementing an interferometer with v\symbol{126}0 at the pi
pulse was published as arXiv:1907.04403v1 and maybe this point can be
discussed by the author.

- refs. [3] and [4] are actually the same: [4] is the arXiv copy of [3] so,
probably, [4] should be removed.

----------------------------------------------------------------------

Report of the Second Referee -- AG11947/Dubetsky

----------------------------------------------------------------------

"Optimization of the atom interferometer phase produced by the set of
cylindrical source masses to measure the Newtonian gravity constant", by B.
Dubetsky, discusses a geometry for the determination of G. The proposed
geometry is a variant on the one used in 2014 by G. Tino and coworkers. The
article seems reasonable, careful, and correct. However, the extremely
technical nature of the manuscript and its limited target audience make it
inappropriate for Physical Review A. I recommend that the author consider a
more specialized journal for the work.

\subsection{Date: Sunday, November 17, 2019 8:05:22 PM}

From: Boris Dubetsky

To: pra@aps.org

Subject: Re: Your\_manuscript AG11947 Dubetsky

Attachments: cylinders5\_short.pdf

Dear Dr. Yin,

I have studied carefully the Referees' reports and modified the manuscript
along with their recommendations. Please, see below my answers to the
Referees conclusions and suggestions, where I also listed the modifications.

I disagree with Referees' conclusions and ask you to consider the attached
new version of the manuscript for publication.

Yours sincerely,

Boris Dubetsky

----------------------------------------------------------------------

Answer on the Report of the First Referee

----------------------------------------------------------------------

1. Referee concluded: "I think that the material is potentially interesting
but in its present form it cannot be considered for publication."

I've changed the form and content of the manuscript along with Referee
recommendations. I hope after those changes Referee would change his opinion.

2. Referee recommended writing supplementary contents.

I wrote the Supplemental Material and replaced reference on the online
version of the article with the reference on the Supplemental Material.

3. I am appreciated to the Referee findings of the previous expressions of
the cylinder gravity field derived by Drs. Miles and Nabighian. In contrast
to my calculations, Drs. David Miles and N.Nabighian do not use the
technique proposed in the textbook [20], and, therefore, in my opinion,
previous derivations lengthier. I wrote the code for the Drs. Miles and
Nabighian expressions and found numerically that previous expressions do not
coincide with my expressions. Moreover, I tested numerically previous and my
expressions for several limiting cases (large distance, an axial field
component at the plane z=h/2 and at x=y=0) and found that only my
expressions converge to the expected results.

I wrote following email to Dr. Miles:

"Dear Dr. Miles:

Regarding your really interesting article Journal of Applied Geophysics 159
(2018) 621--630, I would like to bring to your attention my recent preprint
arXiv:1907.03352v4 [physics.atom-ph]. In the appendix of the preprint, I
also derived the cylinder gravitational field. I compared numerically your
expression (21) and my expression (A22). To write code for your expression I
used Eqs. (13,17-19,22-25,A-9,A-10,A-11) from your article. I also tested
successfully my Eqs. (A17, A22) for the large distance, where cylinder can
be considered as a point source of gravity.

Unfortunately, your expression (21) and my expression (A22) do not coincide
numerically.

I'd be really appreciated it if you help to resolve this discrepancy.

With best regards, Boris"

I've got no answer from Dr. Miles.

I've added references found by Referee (see references [21,22]), the
previous text

"The technique for calculating the gravitational field without calculating
the gravitational potential was proposed in book [20], but the final
expression for the cylinder field is given in [20] without derivation.
Following technique [20], we calculated the field and arrived at expressions
(A17, A22) that do not coincide with those given in [20]. Both the
derivations and final results are presented in this article."

was replaced on page 2 with the following text

" Expressions for the field of the cylinders have been derived in the
articles [21,22]. Alternatively, the technique for calculating the
gravitational field without calculating the gravitational potential was
proposed in the book [20], but the final expression for the cylinder field
is given in [20] without derivation. Following technique [20], we calculated
the field and arrived at expressions (A17, A22). Our expressions do not
coincide with those given in [20-22]. Both the derivations and final results
are presented in this article. Following the derivations in the articles
[21,22], we are going to find out analytically the reason of the
discrepancies between different expressions and publish it elsewhere".

4. Referee disagree "that a shift of the order of -200 ppm should be applied
to the value of G published in [3]". It is because the use of a Montecarlo
did not bring to any shift. Montecarlo is just an alternative method of
generating the error budget. Evidently, two different methods, Montecarlo
and expectation values' calculations in my manuscript should bring to the
same error budget. I failed to find in [3] the information necessary to
verify the Montecarlo in [3], while all details of my calculations are
presented in the Eqs. (10, 17) , Table S.I, which brought me to the Eqs.
(S.1), which in turn for Standard deviations (24), which I took from [3],
lead to the shift (25b) and it changes the gravity constant G to the value
(51).

Referee agrees with me that the use of maximal or minimal points leads to
the "substantial bias". I've calculated this bias (200 ppm). I did not get
from [3] a piece of information about bias in the Montecarlo. Sorry but,
logically, if Referee agrees with a bias then he should agree with the
consequence of the bias, i. e. with the shift of G.

In addition, I would like to pay Referee attention to the phrase in [3]:
"Extracting the value of G from the data involved the following steps:
calculation of the gravitational potential produced by the source masses;
calculation of the phase shift for single-atom trajectories; Monte Carlo
simulation of the atomic cloud; and calculation of the corrections for the
effects not included in the Monte Carlo simulation (Table 1)." Uncertainties
to the atomic cloud position are included in the Table 1. Then from the
cited text, I can conclude that uncertainties in atomic position (and,
maybe, velocity) were not included in the Montecarlo simulation in [3].

Regarding this point I added the following text on page 11:

"We would like to note that Monte Carlo simulation in [3, 4] did not bring
to any shift caused by uncertainties of the atom position and velocity."

5. Referee wrote: "Some choices of numerical parameters seem peculiar. Can
for example the author motivate the choice of "a distance between the floors
of 0.20011 m" (why the extra 110 um?) or "temperature 115 pK and radius of
the atom cloud 170 um (which are larger than those observed in [7])" (why
then not use then the numbers given in [7] in an estimate)." I had the
following motivations:

- distance between the floors of 0.20011 m

I performed calculations for the cylinders chosen in [3,4], they have a
peculiar height h=15.011cm. I added non-peculiar distance dh=5cm between
cylinders tops and bottoms (see Fig. 4a) and got the distance between floors
of 0.20011 m. Regarding this choice, I added a sentence on page 9 "Distance
between floors includes the height of the cylinder h = 0.15011m [3, 4] and
distance dh = 5cm between cylinders tops and bottoms."

- temperature 115 pK and radius of the atom cloud 170 um

These parameters were not chosen for the calculations in the manuscript. I
showed in [5] that at these values of parameters one can in principle
measure G with an accuracy 200ppb. I referred to the article [7] only to
justify that parameters required for 200 ppb are achievable because even
better parameters have been observed in [7]. To underline this point I added
a sentence on page 1 "For the parameters achieved in [2-4] at present, the
optimal preparation of the atomic clouds and proper positioning of the
gravity sources can also lead to an increase in the accuracy of the
G-measurement"

6. Referee wrote " but I would recommend to simplify some formulas in order
to improve readability. As an example consider eq.(20) and eq. (21a-e)".

I followed this recommendation, see Eqs. (20 - 23e)

7. I introduced recoil velocity v\_r, see Eq. (22).

8. Referee wrote "By the way, to erase the Earth's gradient k should change
at every pulse. Which k should be taken here? "

To clarify I added on the page 2 following text:

"Since the magnitude of the gravity-gradient tensor $\Gamma $\_\{E\} is
small, the change in the effective wave vector in (3) can be considered as a
perturbation. Another small perturbation here is the gravity field of the
source mass [10]. Since we are not going to consider the simultaneous action
of these two perturbations, we can calculate the parts of the phases $%
\varphi $\_\{s\}\symbol{94}\{(C,F)\} assuming that all three Raman pulses
have the same unperturbed effective wave vector k".

9. I replaced \TEXTsymbol{\backslash}gamma\_zm with \TEXTsymbol{\backslash}%
Gamma\_s3m, \TEXTsymbol{\backslash}chi\_zmn with \TEXTsymbol{\backslash}%
chi\_s3mn and g\_z with g\_3

10. Subscript "2" is used for the second atom cloud so that instead of
replacing \TEXTsymbol{\backslash}Delta\symbol{94}2 with \TEXTsymbol{%
\backslash}Delta\_2, I replaced \TEXTsymbol{\backslash}Delta\symbol{94}2
with \TEXTsymbol{\backslash}Delta\symbol{94}(2)

11. I added text "We assume that variables \{q$_{1}$,\ldots q\_\{n\}\} are
statistically independent."

12. I replaced x with q in Eqs. (15, 16) and wrote "q\_\{i\} is either a
position x\_\{i\} or a velocity v\_\{i\},"

13. Yellow color for atomic trajectories was chosen in [3,4]. But,
nevertheless, I changed it to the red color now.

14. I am appreciated to the Referee for the point that v\_z should not be
equal 0 at the time of pulse action. I was not familiar with this
restriction. Fortunately, the small value of v\_z at the second pulse
action, v\_z \symbol{126}-6mm/s, leads to the detuning 2kv\_z \symbol{126}%
-2.e5/s from Raman resonance for not desirable effective wave vector -k. If
Pi-pulse has duration \TEXTsymbol{\backslash}tau \symbol{126}60us, then the
detuning is an order of magnitude larger than 1/\TEXTsymbol{\backslash}tau,
then the probability of not desirable excitation with recoil -v\_r is 
\symbol{126}4.e-2. In addition, v\_z \symbol{126}-6mm/s is twice larger than
the thermal velocity in [3,4], 3mm/s, i.e. portion of atoms with velocity
v\_z \symbol{126}6mm/s is exponentially small, \TEXTsymbol{<}1.e-8.
Regarding this point I added the following text on page 7:

"Repeating iterations (31) three times we determined the velocities
v\_\{zmax\} and v\_\{zmin\} with an accuracy of 10$^{-7}$m/s. These
velocities match up to the 5th digit. They are also close to the velocity of
the atomic fountain [31] v=gT. differing from it only in the third digit,

$\delta $v=v\_\{zmax\}-gT$\approx -6\cdot 10^{-3}$m/s \ \ \ \ \ \ \ \ \ \ \
\ \ \ \ \ \ \ \ \ \ \ \ \ \ \ \ (32).

This difference, however, is sufficient to exclude the parasitic signal
[32], which occurs when atoms interact with a Raman pulse having an opposite
sign of the effective wave vector. Indeed, the Raman frequency detuning for
the parasitic signal $\delta =2k\delta v\approx -2\cdot 10^{5}s^{-1}$. If
the duration of the $\pi $-pulse $\tau \sim $60$\mu $s, then the absolute
value of the detuning $\delta $ is an order of magnitude greater than the
inverse pulse duration, and the probability of excitation of atoms by a
parasitic Raman field is negligible, is estimated to be about 4\%.

In addition, the velocity (32) is twice as large as the thermal velocity in
the atomic cloud, v=$3\cdot 10^{-3}$m/s [3,4], and therefore the portion of
atoms having the opposite velocity $-\delta v$ and being in resonance with
the parasitic Raman pulse is also exponentially small, is no more than 10$%
^{-8}$".

15. Referee recommends that ref. [4] should be removed. Unfortunately, at
the important for me point, refs [3] and [4] do not coincide, ref. [3] says
"The central pi pulse occurs about 6 ms after the atoms reach the apogees of
their trajectories", while the ref. [4] says "The central pi pulse occurs
about 6 ms before the atoms reach the apogees of their trajectories". Ref.
[4] has been published later than ref. [3]. That is why to get the time of
apogee, and launching velocity, I used a sentence from ref. [4] and inserted
ref. [4] in the manuscript.

----------------------------------------------------------------------

Answer on the Report of the Second Referee

----------------------------------------------------------------------

Second Referee negative conclusion based on "the extremely technical nature
of the manuscript and its limited target audience."

It is pretty common that gravitational constant is measured much worse than
any other fundamental constants. My manuscript proposes a new technique to
measure this constant, and, therefore, it could be interesting to some
degree wide audience. I found for the first time that the error budget has
to be revised and developed the revision. This point also could be
interesting to some degree wide audience.

To minimize technical things I took out all tables and lengthy equations and
put them into Supplemental material. But, evidently, if I wrote about
revision then I have to demonstrate the revision consequences.

I am sorry, but the Report of the Second Referee does not contain any
specific points, and, therefore, I did not make any changes following this
Report.

\subsection{Date: Friday, November 22, 2019 10:28:37 AM}

From: pra@aps.org

To: bdubetsky@gmail.com

Subject: Your\_manuscript AG11947 Dubetsky

Re: AG11947

Optimization of the atom interferometer phase produced by the set of
cylindrical source masses to measure the Newtonian gravity constant by B.
Dubetsky

Dear Dr. Dubetsky,

The above manuscript has been reviewed by two of our referees. Comments from
the reports appear below. Although the referees gave conflicting advice, we
agree with the assessment of the second referee. We feel that the manuscript
is a very specialized study that may not be of general interest to our broad
audience in the AMO community.

We regret that in view of these comments we cannot accept the paper for
publication in the Physical Review.

Yours sincerely,

Xiangyu Yin

Assistant Editor

----------------------------------------------------------------------

Second Report of the First Referee -- AG11947/Dubetsky

----------------------------------------------------------------------

After the author's revision, the manuscript is self contained so I am in
favor of publication in Phys. Rev. A.

Keeping the original numbering in the author's response I have some comments
only on the two following points:

4. On this point I feel that the author and I do agree on the heart of the
matter but there is a misunderstanding on a crucial point. I apologize
therefore with the editor and the author for entering, in the following, in
a detailed trivial example to try to clarify the issue.

Take a standard gaussian variable x \symbol{126}N(0,1) and consider y=x%
\symbol{94}2. We actually know the distribution of y i.e. a chisquare with
one degree of freedom. Of course estimating E[y] as (E[x])\symbol{94}2 is a
bad idea since we can be sure that E[y]\TEXTsymbol{>}0. Actually we know
that E[y]=1.

We could obtain a better estimate of E[y] either using E[x]=0 and adding the
shift given in eq. (17b) or by running a Montecarlo simulation, generating
say 10\symbol{94}5 normally distributed variables x\_i computing y\_i=x\_i%
\symbol{94}2 and taking the average of all the y\_i. Note that in the second
case I do not even need to know E[x]. The average of the y\_i is an unbiased
estimate of E[y].

The author writes: "I did not get from [3] a piece of information about bias
in the Montecarlo...".

As I pointed out above, it's because, as I understand it, it's not usually
provided since it is not useful to state the result as E[y]=(E[x])\symbol{94}%
2+bias since the Montecarlo provides directly E[y].

In an experimental situation however the a-priori distribution of x is not
known so x \symbol{126}N(0,1) should be understood as: a normal distribution
is assumed and mean(x)=0 and var(x)=1 are actually experimental values with
an estimated or measured uncertainty.

As an example let's assume var(x)=1.0+- 0.1.

Running a single Montecarlo is therefore not enough and I could try to see
what happens to E[y] when picking for var(x) [0.9, 1.0, 1.1] for a simple
example. From the three values for E[y], I obtain estimates

both for bias and uncertainty for E[y] due to the uncertainty on var(x).
Here actually, since we know that E[y]=var(x), there will be no bias but an
estimate for E[y] of 1.0+-0.1.

In the spirit of this somehow futile example I would add to Table 1 in [3] a
row that reads

var(x) 0.1 -- 0.1

meaning that the uncertainty of 0.1 on var(x) will not add a bias to the
result of the Montecarlo but will contribute with 0.1 to the uncertainty of
the results. Of course with many variables I should be very careful in
assuming statistical independence and adding the uncertainties in quadrature
but this is another matter.

At this point I might have succeeded or failed in convincing the author that
the correction that he evaluates for the value of G in [3] is probably
already in place, but this has no effect on the validity of the manuscript.
The author can choose to keep or remove the claim that the correction should
be applied, according to his judgment.

15. "..refs [3] and [4] do not coincide, ref. [3] says "The central pi pulse
occurs about 6 ms after the atoms reach the apogees of their trajectories",
while the ref. [4] says "The central pi pulse occurs about 6 ms before the
atoms reach the apogees of their trajectories". Ref. [4] has been published
later than ref. [3]. That is why to get the time of apogee, and launching
velocity, I used a sentence from ref. [4] and inserted ref. [4] in the
manuscript."

Fair enough. However I would suggest that, for better clarity, instead of
always citing [3,4] together, without explaining the difference mentioned
above, a difference that will be easily missed even by the most dedicated
reader of the article, that [3] is cited and a footnote is added explaining
the discrepancy between [3] and its successive arXiv copy, [4], while
stating that the latest published value was used.

-------

A misprint at the end of sec. II: "For the each case considered.."

----------------------------------------------------------------------

Second Report of the Second Referee -- AG11947/Dubetsky

----------------------------------------------------------------------

In my original report, I found this manuscript to be too narrow in its
audience to be appropriate for Physical Review A. My opinion is unchanged by
the iterative changes made to the manuscript since its first submission.

The references cited give strong support to my assessment. The majority of
the 32 references fall in four categories:

1, Citations to work by two experimental groups, the Kasevich group and the
Tino group. [10 citations]

2, Self-citations to the author's previous work [7 citations]

3, Papers published 20 or more years ago and textbooks [7 references]

4, Unpublished work [6 citations]

I also note that there are only two PRA papers, one of which is from 1986.
In the introductory paragraph of this manuscript, where one normally
discusses the broad appeal of a scientific question, only papers by the
author himself, and the two experimental groups mentioned above, are cited.
(ie, groups 1 and 2 above.)

To me, this is an indication that the topic is no longer of current or broad
interest to the PRA community. I do not say that the manuscript is incorrect
or misleading, but rather suggest the author submit it to Metrologia, an
appropriate journal (of comparable impact factor as PRA) that is dedicated
to the kind of detailed analysis presented here. Citations 13 and 14 are
from that journal, for instance.

\subsection{Date: Friday, November 22, 2019 2:23:17 PM}

From: Boris Dubetsky

To: pra@aps.org

Subject: Re: Your\_manuscript AG11947 Dubetsky

Dear Dr. Yin,

In my answer to the First Report of the Second Referee, to prove that the
manuscript could be of common interest, I wrote that it is devoted to the
measurement of the gravitational constant, i.e. to the topic which does have
a wide auditorium. I would like to pay your attention that Second Referee
totally ignored this my argument. Instead, Second Referee performed an
analysis of the bibliography. I do not know why the bibliography weights
more than the topic of the manuscript. I thought that the elaboration of new
methods and revision of the inappropriate approach inside important topic
should overcome a bibliography.

But even regarding bibliography, I do not know why Kasevich group, Tino
group, Steven Chu group, A. Roura sent their articles to the non-specialized
journals Physical Review Letters, Science and Nature, and did not publish
those articles in the Physical Review A. But, I am sure that articles
published in those journals are of a great interest for the Physical Review
A community. I mean 10 References [2,3,6-8,11,12,23,29]. Together with
references [9, 31], about which Second Referee wrote, we have 12 references,
which are definitely interested in the Physical Review A community. Five
articles [3,7-9,12] are less than 5 years old. This brief study shows that
from the bibliography point of view my manuscript could be also not rejected.

Would you please to reconsider your conclusion about the manuscript.

Sincerely yours,

Boris Dubetsky

\subsection{Date: Friday, November 22, 2019 3:40:39 PM}

From: pra@aps.org

To: bdubetsky@gmail.com

Subject: AG11947 Dubetsky

Re: AG11947

Optimization of the atom interferometer phase produced by the set of
cylindrical source masses to measure the Newtonian gravity constant by B.
Dubetsky

Dear Dr. Dubetsky,

Your paper has been rejected. Further consideration can only be given if you
decide to exercise the option, available under this journal's Editorial
Policies (copy attached), of appealing the decision to reject the
manuscript. Adjudication of such an appeal is based on the version of the
manuscript that was rejected; no revisions can be introduced at this stage.

Yours sincerely,

Xiangyu Yin

Assistant Editor

\subsection{Date: Wednesday, January 22, 2020 1:45:22 PM}

From: pra@aps.org

To: bdubetsky@gmail.com

Subject: Your\_manuscript AG11947 Dubetsky

Attachments: ag11947\_report\_eb.pdf

Re: AG11947

Optimization of the atom interferometer phase produced by the set of

cylindrical source masses to measure the Newtonian gravity constant

by B. Dubetsky

Dear Dr. Dubetsky,

Your formal appeal of this manuscript has been evaluated by an Editorial
Board Member. The EBM advises us not to publish in Physical Review A, and we
accept this advice. Your appeal has been considered,

and our decision to reject is maintained. This concludes the scientific
review of your manuscript.

Yours sincerely,

Xiangyu Yin

Assistant Editor

\subsubsection{Report \TEXTsymbol{\vert} Editorial Board}

Re: AG11947

Optimization of the atom interferometer phase produced by the set of
cylindrical source masses to measure the Newtonian gravity constant

by B. Dubetsky

The manuscript written by B. Dubetsky deals with the calculation of the
gravitational field of a homogeneous cylinder and the measurement of the
Newtonian gravitational constant G by atomic interferometry methods (phase
estimates). The author uses a technique that aims at eliminating Earth's
gravity-gradient. The main claim is that higher order terms in the expansion
tend to be important and cannot be neglected, and that textbook (Chen, Cook,
Ref. [20]) calculations of the gravity field of the cylinder do not take
important factors into account. The author also claims that there are
discrepancies among his calculation and those performed in Refs. [21,22].
One of the main results of the ms is Eq. (54), where a different value is
obtained for the gravitational constant (as compared to that reported in
Refs. [3,4]).

The two referees offered conflicting advice. Referee 1 made a number of
important technical observations, and is clearly very interested in the
topic. After the second round of refereeing, she/he concluded that the
manuscript can be accepted for publication. On the other hand, Referee 2
wrote that "the extremely technical nature of the manuscript and its limited
target audience make it inappropriate for Physical Review A." Referee 2 also
suggested a possible alternative venue (Metrologia), where the ms can be
submitted.

Although the opinions of both referees are sensible, I agree with the
standpoint of Referee 2. The manuscript is very technical and difficult to
read. The calculation can be of interest for a limited audience, such as
experimental groups working on this topic (Tino's and possibly Kasevich's)
and possibly some other researchers involved in similar analyses. It is not
suitable for a wider readership, such as that of Physical Review A.
Moreover, as one Editor wrote, an article submitted to Physical Review A
should be well organized, and clearly written in scientific English. This
manuscript is unfortunately far from meeting these criteria.

Three Editors were involved with the present manuscript: Associate Editor of
Physical Review A Doerte Blume, Managing Editor of Physical Review Research
Juan-Jose Lietor-Santos, and Assistant Editor of Physical Review A Xiangyu
Yin. All of them concluded that the manuscript is not suitable for
publication in Physical Review A. I kindly ask Dr. Dubetsky to consider that
the Editors have evaluated this case with care and their consolidated
editorial expertise.

In conclusion, I uphold the rejection of the Editors of Physical Review A
and Physical Review Research and, regretfully, do not recommend publication
in Physical Review A.

Saverio Pascazio

Editorial Board

Physical Review A
\end{subequations}

\end{document}